\documentstyle[aps,pra,psfig]{revtex}

\begin{document}
\def \blank{\mbox{}}

\def\d{\mbox{$\partial$}}
\def\Pop{\mbox{\sf P}}
\def\de{\mbox{$d_E$}}
\def\dw{\mbox{$d_W$}}
\def\dl{\mbox{$d_L$}}
\def\da{\mbox{$d_A$}}
\def\hb{\mbox{$\hbar$}}
\def\Re{\mbox{\,Re\,}}
\def\Im{\mbox{\,Im\,}}
\def\dw{\mbox{$d_W$}}
\def\nb{\mbox{$N_B$}}
\def\A{\mbox{\bf A}}
\def\B{\mbox{\bf B}}
\def\C{\mbox{\bf C}}
\def\M{\mbox{\bf M}}
\def\P{\mbox{\bf P}}
\def\x{\mbox{{\bf x}}}
\def\ol{\mbox{$\bar{\lambda}$}}
\def\Q{\mbox{{\bf Q}}}
\def\R{\mbox{{\bf R}}}
\def\S{\mbox{{\bf S}}}
\def\L{\mbox{{\bf L}}}
\def\N{\mbox{{\bf N}}}

\def\Rd{\mbox{$R^d$}}
\def\DF{\mbox{{\bf DF}}}
\def\DG{\mbox{{\bf DG}}}
\def\DH{\mbox{{\bf DH}}}
\def\DFL{\mbox{$\DF^L$}}
\def\DGL{\mbox{$\DG^L$}}
\def\OSL{\mbox{{\bf OSL}}}
\def\T{\mbox{$T$}}
\def\bfE{\mbox{${\bf E}$}}
\def\Ec{\mbox{{$\cal E$}}}
\def\Nc{\mbox{{$\cal N$}}}
\def\Pc{\mbox{{$\cal P$}}}
\def\Ac{\mbox{{$\cal A$}}}
\def\l{\mbox{$\lambda$}}
\def\g{\mbox{$\gamma$}}
\def\oma{\mbox{$\omega_A$}}
\def\om{\mbox{$\omega$}}
\def\eps{\mbox{$\epsilon$}}
\def\Y{\mbox{$\bf{Y}$}}
\def\y{\mbox{$\bf{y}$}}
\def\v{\mbox{$\bf{y}$}}
\def\ynn{\mbox{$\bf{y}^{NN}$}}
\def\w{\mbox{$\omega$}}
\def\z{\mbox{$\bf{z}$}}
\def\e{\mbox{$\epsilon$}}
\def\del{ \mbox{\boldmath{$\Delta$}} }
\def\delt{ \mbox{\boldmath{$\delta$}} }
\def\xint{\mbox{$\displaystyle\int d^3x \,$}}
\def\f{\mbox{$\bf{f}$}}
\def\p{\mbox{$\bf{\rho}$}}
\def\F{\mbox{$\bf{F}$}}
\def\E{\mbox{$\bf{E}$}}
\def\H{\mbox{$\bf{H}$}}
\def\G{\mbox{$\bf{G}$}}
\def\U{\mbox{$\bf{U}$}}
\def\J{\mbox{$\bf{J}$}}
\def\doublespace{\parskip 4pt plus 1.5pt
                \baselineskip 24pt plus 1pt minus .5pt
                \lineskip 6pt plus 1pt \lineskiplimit 5pt}

\newcommand{\be}{\begin{equation}}
\newcommand{\bea}{\begin{eqnarray}}
\newcommand{\ee}{\end{equation}}
\newcommand{\eea}{\end{eqnarray}}

%
%

%
%
\newcommand{\av}[1]{{\langle #1 \rangle}}

%
%
\newcommand{\avabs}[1]{{\langle |#1| \rangle}}

%
%
\newcommand{\avabssq}[1]{{\langle |#1|^2 \rangle}}

\title{Synchronization of chaotic oscillations in doped fiber ring lasers}

\author{Clifford Tureman Lewis\cite{mikecliffbyline},
Henry D. I. Abarbanel\cite{hdiabyline},
Matthew B. Kennel\cite{mbkbyline},
Michael Buhl\cite{mikecliffbyline},
and Lucas Illing\cite{mikecliffbyline}}
\address{
Institute for Nonlinear Science\\
University of California, San Diego\\
La Jolla, CA 92093-0402}

\date{\today}

\maketitle

\begin{abstract}

We investigate synchronization of and subsequently communication using
chaotic rare-earth-doped fiber ring lasers, represented by a physically
realistic model introduced in~\cite{akbl1}. The lasers are first coupled by
transmitting a fraction $c$ of the circulating electric field $\E_T$
in the transmitter and injecting it into the optical cavity of the
receiver, e.g., by fiber optics.  We then analyze a coupling strategy which relies on modulation of the intensity of the light alone. This avoids problems associated with polarization and phase of the laser light.

We study synchronization as a
function of the coupling strength and see excellent convergence, even
with small coupling constants.  We prove that in an open loop
configuration ($c=1$) synchronization is guaranteed due to the particular
structure of our equations and of the injection method we use
for these coupled laser systems~\cite{prlak}.

We also analyze the generalized synchronization of these model lasers
when there is parameter mismatch between the transmitter and the
receiver. The synchronization is insensitive to a wide range of
mismatch in laser parameters, including the pumping level and the gain
in the active medium, but it is sensitive to other parameters, in
particular those associated with the phase and the polarization of the
circulating electric field, and the lengths of the passive fiber
rings.  Less restrictive synchronization metrics reduce the
sensitivity to mismatch. 

We then address communicating information between the transmitter and
receiver lasers. We investigate a scheme for modulating information
onto the chaotic electric field and demodulation and detection of the
information embedded in the chaotic signal passing down the
communications channel.  The modulation scheme consists of an
invertible operation in which the message to be transmitted, raised
from its baseband status to the optical passband, is combined with the
light in the transmitter. At the receiver, we invert the combining
operation and recover the message.

Many of the ideas we consider have been demonstrated experimentally
by Roy and VanWiggeren~\cite{science,prlrvw,private}. Their remarkable
experimental success at high symbol rates and low bit error rates attests
to the viability of the communications methods we investigate here.

Finally we comment on the {\it cryptographic setting} of our algorithms,
especially the open loop strategy at $c=1$, and hope this may lead others
to perform the cryptographic analyses to determine which, if any, of the communications
strategies we investigate are secure. At $c=1$ our methods are known in the cryptographic literature as self-synchronous stream ciphers operating in cipher feedback mode. For $c \ne 1$, ours is a type of cryptographic system apparently not known in that field. Neither a cryptanalysis nor any claims for security of
communication are made here.

\end{abstract}

\tableofcontents

\section{Introduction}

Synchronization of chaotic oscillators is a phenomenon found quite
often in physical and biological systems. The idea that chaotic
systems could synchronize their motions was suggested some time ago by
Fujisaka and Yamada~\cite{fuji} and independently by Afraimovich,
Rabinovich and Verichev~\cite{avr}. An early investigation of
synchronization in neural networks~\cite{hansel} explored application
in a wider arena.

The idea was again independently proposed and
then experimentally explored by Pecora and Carroll~\cite{pc1}. The latter
authors also suggested that the use of synchronized chaotic oscillators for
communications would be of some interest. The work of Pecora and
Carroll led to the investigation of a wide variety of synchronized
chaotic systems~\cite{special} including close relatives of those we
discuss in this paper.

\begin{figure}
\centerline{\psfig{file=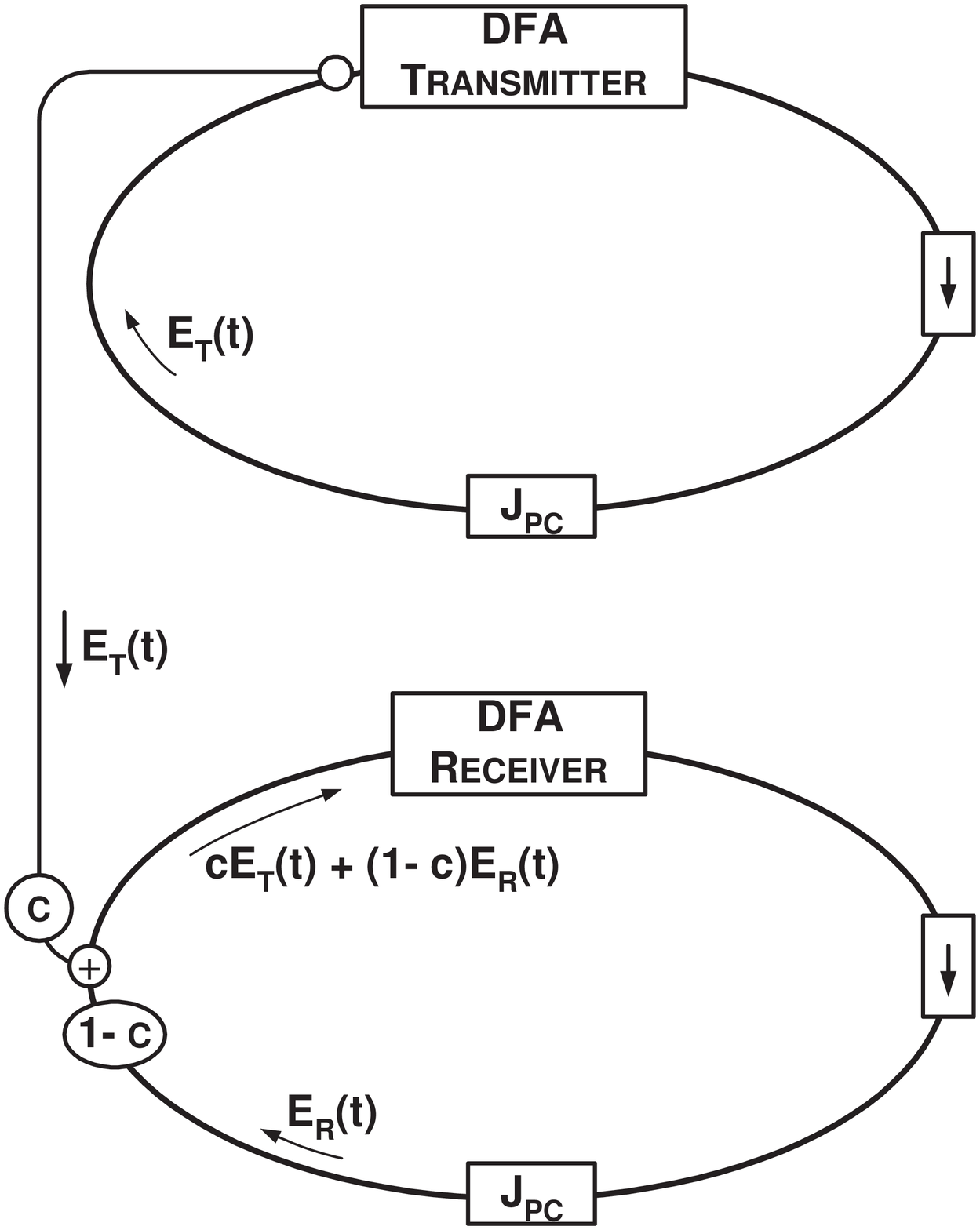,width=3in}}
\caption{The setup of coupled DFRLs. The electric field
circulating in the transmitter laser is $\Ec_T(t)$. After it is transmitted
through a channel to the receiver, a fraction $c\Ec_T(t)$ is injected into
the input of the rare earth doped amplifier in the receiver ring. At the
same time a fraction $(1-c)\Ec_R(t)$ of the field circulating in the receiver
ring is added to it, so the net field injected into the amplifier input is
$c\Ec_T(t) + (1-c)\Ec_R(t)$.}
\label{fig:scheme}
\end{figure}

In this paper we explore the synchronization properties of the models
we have built for rare-earth-doped fiber ring lasers
(DFRLs)~\cite{akbl1}. Our plan here is to use two such model lasers
connecting them by transmitting the electric field $\E_T$ circulating
in one laser (the transmitter) to a second identical laser (the
receiver). Into the optical cavity of the receiver laser we re-inject
a fraction $(1-c)\E_R$ of the receiver field $\E_R$ and also add a
fraction $c \E_T$ of the field arriving from the first laser. The net
field entering the receiver laser is then $c\E_T + (1-c)\E_R$. This
setup is sketched in Fig. \ref{fig:scheme}. When the lasers are
synchronized so that $\E_R = \E_T$, then the combination $c\E_T +
(1-c)\E_R$ is independent of c and equal to the field in either
laser. As we vary c, we change the precise combination of transmitter
field and receiver field which is seen at the receiver.

We investigate the synchronization both for identical transmitter and
receiver, and then for lasers which have various parameter mismatches 
including the gain and pumping of the active medium,
their polarization characteristics, and their ring
length. The synchronization is quite robust for mismatches in gain and
pumping power, but it is very sensitive to mismatches in polarization
or phase characteristics of the transmitter and receiver. This is
shown in some detail, and the origin of the loss of synchronization is
analyzed. 

Because of this sensitivity, we investigate another coupling
strategy which uses only the intensity of the circulating electric
fields to connect the transmitter and receiver. This is shown to be
a potentially viable method of synchronizing the lasers, and an example of
communicating using amplitude modulation of the transmitter intensity
is investigated.

Many of the papers on synchronization of chaotic systems have dealt with
applications to communications. While we are focused here on the use of
doped fiber ring lasers, the principles associated with synchronization and
communication are shared by earlier investigations, in particular at $c=1$
by the methods of Rulkov and Volkovskii~\cite{rv} and those of Kocarev
and Parlitz~\cite{parlitz}.

The attraction of using chaotic signals as the carriers of information
comes from their potentially efficient use of channel bandwidth as
well as the low cost and power efficiency of some chaotic transmitters
and receivers~\cite{sush}. There has also been much discussion of the
security of communications systems based on chaotic carriers. The
security of any such system depends in detail of the modulation
methods used, not on the perception that the carrier is complex or
temporally irregular. Whether any given modulation/demodulation
strategy is secure needs to be determined by a rigorous cryptographic
analysis, and in this paper we do not discuss this issue or attempt to
provide such an analysis.  We do comment in this paper on the {\it
cryptographic setting} of our methods at $c =1$ where similar
algorithmic structures using linear shift registers have been known
and subjected to cryptographic scrutiny. When the transmitter and
receiver are synchronized, but the receiver is not run open loop ($c \ne
1$), we know of no systematic analysis for such systems in the same
general cryptographic framework.  This may be a new class of systems.
We hope that this paper and others in this area will stimulate
research into these issues and provide guidelines for design of secure
chaotic communications systems, when they are desired.

\section{Equations for an Individual DFRL}

\begin{figure}
\centerline{\psfig{file=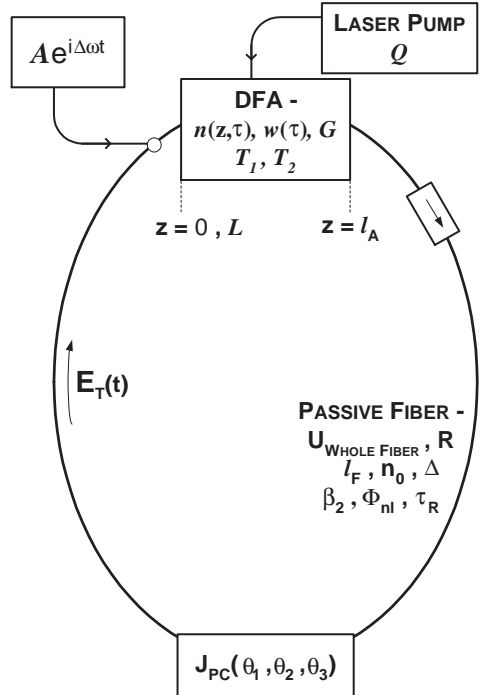,width=2.5in}}
\caption{A schematic diagram showing the relevant location of
the parameters we consider in our model of coupled 
rare-earth-doped fiber lasers.}
\label{fig:diagram}
\end{figure}

We use the model introduced in \cite{akbl1}.  The doped fiber ring
laser (DFRL) we consider contains an optical amplifier composed of 
rare-earth-doped single-mode fiber of length $l_A$, whose active atoms are
pumped by an external laser diode source. Connecting the output of
this active section to its input is a piece of passive fiber of length
$l_F$. In the passive fiber is an isolator which guarantees the
direction of flow of light, a polarization controller, and a location
where external light of given amplitude and frequency $\omega_I$ can
be injected into the fiber. The total length of the fiber cavity is $L
= l_A + l_F$. The general locations of the relevant parameters for each laser 
is shown schematically in Fig. \ref{fig:diagram}.

The electric fields $\E(z,t)$ are described by the envelope $\Ec(z,t)$ of
the optical plane wave of frequency $\omega_0$, $\E(z,t) = \Ec(z,t)e^{i(k_0
z -\omega_0t)}$ where $\omega_0 = k_0c/n$. The dynamics of $\Ec(z,t)$
consists 
\begin{itemize}
\item of linear birefringence of magnitude 
\be
\Delta = n_0 (n_x-n_y),
\ee
where $n_x$ and $n_y$ are the indices of refraction along the principal
axes of the fiber, and $n_0 = (n_x+n_y)/2$, 
\item of group velocity dispersion (GVD) which comes from second order
variations in frequency of the linear dispersion relation of the fiber, 
\item of contributions to the polarization of the medium associated with
the population inversion of the atomic levels, and 
\item  of nonlinear polarization effects associated with the Kerr term cubic
in electric field strength.
\end{itemize}

With these effects included, our equations for the propagation of
the electric field envelope $\Ec = (\Ec_x,\Ec_y)$ in the active medium $0
\le z \le l_A$
become, in retarded coordinates $z = z, \tau = t - z/v_g$
with $v_g$ is the group velocity of the waves, 
\begin{equation} \frac{\d\Ec_{x,y}(z,\tau)}{\d
  z} = g n(\tau) \Ec_{x,y} + \L_{x,y} \Ec_{x,y} + \N_{x,y} \Ec_{x,y}.
\label{eq:symbolicprop} 
\end{equation}
$\L$ contains the linear parts of the propagation operator excluding
gain, and $\N$ the Kerr nonlinearity.  The linear operator $\L$,
including birefringence, GVD and gain dispersion, is most naturally
represented in the Fourier domain: 
\begin{equation}
 \L_{x,y} = \pm \frac{i k_0
(n_x-n_y)}{2 n_0} \mp \frac{ \Delta}{n_0c} i\w - \frac{i}{2} \beta_2
\w^2 - \frac{gn(\tau) \w^2 T_2^2}{1 + \w^2 T_2^2}
\end{equation}
with $\w$ the signal angular frequency. The first term in $\L_{x,y}$
only results in an overall arbitrary phase shift for the two
polarizations, which can be absorbed without loss of generality into
the parameterization of the passive section described below. The term
linear in $\w$ represents linear birefringence; the next term,
quadratic in $\w$, is the group velocity dispersion.  The last term is
associated with the gain curve, and arises from the fact that the
center frequency of the line $\w = 0$ is amplified more strongly than
frequencies on either side of the line.  The nonlinear operators are
\begin{eqnarray}
\N_x \Ec_x &=& i \chi_3 \biggl\{ \left( |\Ec_x(z,\tau)|^2 + \frac{2}{3}
|\Ec_y(z,\tau)|^2 \right) \Ec_x(z,\tau) + \frac{1}{3} \Ec_x^*(z,\tau)
\Ec_y(z,\tau)^2\biggr\}
\label{eq:kerroperatorx}  \\ 
\N_y \Ec_y &=& i \chi_3 \biggl \{ \left( |\Ec_y(z,\tau)|^2 + \frac{2}{3}
|\Ec_x(z,\tau)|^2 \right) \Ec_y(z,\tau) + \frac{1}{3} \Ec_y^*(z,\tau)
\Ec_x(z,\tau)^2\biggr \}.
\end{eqnarray}

The physical implication of this third-order nonlinearity is most easily
characterized by the non-dimensional phase shift $\Phi_{nl}$ experienced 
by an $\Ec$ field as it passes through the fiber given by~\cite{agrabook}
\be
\Phi_{nl} = \frac{\chi_3 \pi L}{\lambda}(P_a + 2P_b),
\ee
where $P_a$, $P_b$ are the optical powers in the parallel and perpendicular 
direction. We parameterize our simulations by $\Phi_{nl}$ and vary its value
to tune in a desired amount of optical nonlinearity. Physically this
could be interpreted as changing various parameters such as active medium pumping,
which would increase the optical power, or changing the value of the ring
fiber length L.

These equations must be solved numerically to propagate the light from
its entry into the doped fiber amplifier (DFA) at $z=0$ to its exit at
$z=l_A$, represented here by a propagation operator on the
vector $\Ec = \left[\Ec_x, \Ec_y \right]$:  
$\Ec(z = l_A, t + l_A/v) = \Pop \{\Ec(z=0,t)\}$.  

The polarization of the electric field in the ring is affected by the
birefringence in the fiber arising from numerous small effects
associated with imperfections in the fiber, strains, etc.
Following~\cite{polfiber} we write the net effect of the fiber on the
polarization states of the field as a unitary Jones matrix which we
call $\U_{\mbox{Whole Fiber}}$.

The overall propagation map including all the passive parts of the ring
and external injection is,  
\begin{equation}
\Ec(t + \tau_R) = \Ac(t+\tau_R) e^{i(\omega_I - \omega_0)(t+\tau_R)} +
\biggl( \R \J_{PC} \U_{\mbox{Whole Fiber}}\biggr)
\Pop
\{\Ec(t)\}. \label{eq:fullemap}
\end{equation}
Reading from left to right the terms are external monochromatic injection,
possibly polarization dependent attenuation, $\R = \mbox{diag}(R_x,R_y)$,
the unitary Jones matrix for the polarization controller $\J_{PC}$, the 
matrix for the passive fiber, and the propagator through the active medium.   
This discrete-time map, a recursion relation between the
field at a time $t$ and at a time $\tau_R$ later, is one of the dynamical
rules of our ring laser system. The other is the population inversion
equation in its simplified form.

The physics of the atomic polarization in the active medium is governed
by the usual Bloch equation for the population inversion at time $\tau$
and spatial location z:
\be
  \frac{\d n(z,\tau)}{\d \tau} = Q - \frac{1}{T_1}
(n(z,\tau) + 1) - \xi n(z,\tau) |\Ec(z,\tau)|^2, \label{eq:dndt}
\ee
with $Q$ pumping, $T_1$ the lifetime of the excited state (10 ms for a typical
erbium-doped fiber) and $\xi$ a constant relating to the optical cross section
governing the transition rate between levels. Assuming $g$ real, we can integrate
(\ref{eq:dndt}) by ${l_A}^{-1}
\int_0^{l_A} dz$ to arrive at the dynamics for the population inversion
averaged over the entire active medium $w(\tau) = {l_A}^{-1}
\int_0^{l_A} dz\, n(z,\tau)$
\be
\label{eqn:popinv}
  \frac{dw(\tau)}{d \tau} = Q - \gamma
\left( w(\tau) + 1  + \left(
e^{2 l_A g w(\tau)} -1 \right) |\Ec(z=0,\tau)|^2
\right), \label{eq:dwdt}
\ee
where $\gamma$ is the ratio of round trip time to excited state lifetime,
$\tau_R / T_1$.

It is important to note that, except for the gain term, the magnitude
$|\Ec|^2$ is preserved by our propagation operations including the
nonlinear Kerr phase shifts. After the Kerr operator, the main
component of the gain is applied to the two complex envelopes,
multiplying each by $\exp(g l_A w(\tau))$, and finally, the passive
components of the ring, to produce $\Ec_{x,y}(t+\tau_R)$.

Potentially, there is a third dynamical equation associated with the
evolution of the polarization of the medium through which the electric
field passes. The time constant, usually called $T_2$, for this
process is about a picosecond, thus we have safely eliminated that
equation and used the resulting ``static'' polarization in the
population inversion equation.  This is known as a ``class B'' laser. 

We summarize the equations for use below as a map $\M_{\Ec}(w(\tau),\Ec(\tau))$
of the electric field from
time t to time t+$\tau_R$ and a differential equation for the integrated
population inversion:
\bea
\Ec(\tau + \tau_R) &=& \M_{\Ec}(w(\tau),\Ec(\tau)) \nonumber \\
\frac{d w(\tau)}{d \tau} &=& 
Q - \gamma \biggl \{w(\tau) + 1 + |\Ec(\tau)|^2(e^{G
w(\tau)} -1) \biggr \},
\eea
where the active medium specific overall gain term $G$ is defined 
as $G=2 l_A g$.

The details of our numerical schemes are to be found in our earlier
paper~\cite{akbl1}.  Straightforward integration of the partial
differential equations at a resolution sufficient to capture complex
sub-round trip dynamics as seen in experiment results in an algorithm
which is far too slow.  We implemented a scheme which can integrate a
whole round trip at a time, combined with a buffering method to
process the portion of the linear operator in the Fourier domain.
Even still, computation takes approximately 18 hours on a contemporary
workstation to achieve equilibrium (500,000 round trips) on account of
the very large difference in time scales between the fluorescence
lifetime $T_1 \approx 10$ ms associated with an erbium doped fiber and the time resolution of our simulation
$\delta t \approx 80$ ps, necessary to capture the high frequency
dynamics seen in experiment and thus high-bandwidth communication.

\section{Synchronization of Two DFRLs}

We now construct a transmitter and receiver pair and couple their
electric fields by an optical channel.  First we examine the
synchronization of two identical lasers. Next we investigate the
robustness of this synchronization as the transmitter and receiver are
mismatched in various of the physical parameters of the model. We then
look at the robustness of the synchronization against noise in the
communication channel between transmitter and receiver. This will lead
us into considerations of generalized synchronization of the two
optical oscillators.  Finally, we suggest an alternate coupling scheme
and synchronization method constructed to avoid some of the
experimental problems typically encountered with direct optical
coupling.

The values of the model parameters associated with the experiments of 
Van Wiggeren and Roy~\cite{prlrvw,private} used for the group of simulations
in this paper are listed in table~\ref{tab:parameters}.
The model with these parameters produces chaotic waveforms with a
largest Lyapunov exponent, evaluated with the numerical procedure
of~\cite{akbl1}: $\l_1 = (6.3 \pm 0.3)
\times 10^3/s$.

\subsection{Identical Transmitter and Receiver}

In our study of the synchronization of two EDFRLs we have a transmitter
laser with dynamical variables $\Ec_T(t)$ and $n_T(t)$ and a receiver
laser with $\Ec_R(t)$ and $n_R(t)$. The two lasers are started in
different initial conditions, allowed to run several hundred thousand
round trip times uncoupled to reach their asymptotic state.  Coupling is then
activated. The transmitter's field $\Ec_T(t)$ is then injected into
the receiver laser multiplied by a factor $c$ while the circulating
electric field in the receiver is attenuated by a factor $1-c$, and
both are optically recombined before entering the rare-earth-doped
amplifier section of the laser ring. The nonlinear amplifying
element receives $c\Ec_T(t) + (1-c)\Ec_R(t)$ as its input. When the
transmitter and receiver are synchronized, so $\Ec_T(t) = \Ec_R(t)$,
the linear combination $c\Ec_T(t) + (1-c)\Ec_R(t) = \Ec_R(t) =
\Ec_T(t)$ is a solution to the equations of motion for each
laser. This setup is illustrated in Fig. 1.  This kind of coupling is
experimentally achievable with standard fiber optic equipment. 

The equations for this unidirectional coupling between the chaotic systems are for the {\bf transmitter}
\bea
\Ec_T(\tau + \tau_R) &=& \M_{\Ec}(w_T(\tau),\Ec_T(\tau)) \nonumber \\
\frac{d w_T(\tau)}{d \tau} &=& 
Q - \gamma \biggl \{w_T(\tau) + 1 + |\Ec_T(\tau)|^2(e^{G
w_T(\tau)} -1) \biggr \},
\eea
and for the {\bf receiver}
\bea
\Ec_R(\tau + \tau_R) &=& \M_{\Ec}(w_R(\tau),c\Ec_T(\tau) + (1-c)\Ec_R(\tau)) \nonumber \\
\frac{d w_R(\tau)}{d \tau} &=& 
Q - \gamma \biggl \{w_R(\tau) + 1 + |c\Ec_T(\tau) + (1-c)\Ec_R(\tau)|^2(e^{G
w_R(\tau)} -1) \biggr \}.
\eea

When $c=0$, the lasers are uncoupled and run independently. If all of
the physical parameters in the two laser subsystems are identical, the
electric field in each laser visits the same attractor. $\Ec_T(t)$ and
$\Ec_R(t)$ as well as $n_T(t)$ and $n_R(t)$ are uncorrelated due to
the instabilities in the phase space of the system. As we increase $c$
away from zero, we anticipate that for a certain minimum coupling, the
lasers will asymptotically achieve identical synchronization,
$\Ec_T(t) = \Ec_R(t)$, and $n_T(t) = n_R(t)$, even though the population
inversions are not physically coupled.  To exhibit this we run the two
systems for the order of $10^4$ round trips following their
coupling. First, we ask after what time the quantity
\be
H_E(c,t) = \frac{|\Ec_T(t) - \Ec_R(c,t)|}{\avabs{\Ec_T}_{\tau_R}},
\ee
becomes and remains less than some small number. The denominator is the average of
$\Ec_T(t)$ taken over the round trip previous to coupling and gives a
convenient normalization for the magnitude of the synchronization
error $H_E(c,t)$. An EDFRL possesses the property that after the
effects of initial transients, the laser achieves a constant mean
intensity per round trip even though the waveform is chaotic. In our
simulation, the lasers were run long enough before coupling for the
transients to damp out, so in $H_E(c,t)$ the value of
$\avabs{\Ec_T(t)}_{\tau_R}$ is approximately constant in the transmitter
from round trip to round trip.

\begin{figure}
\centerline{\psfig{file=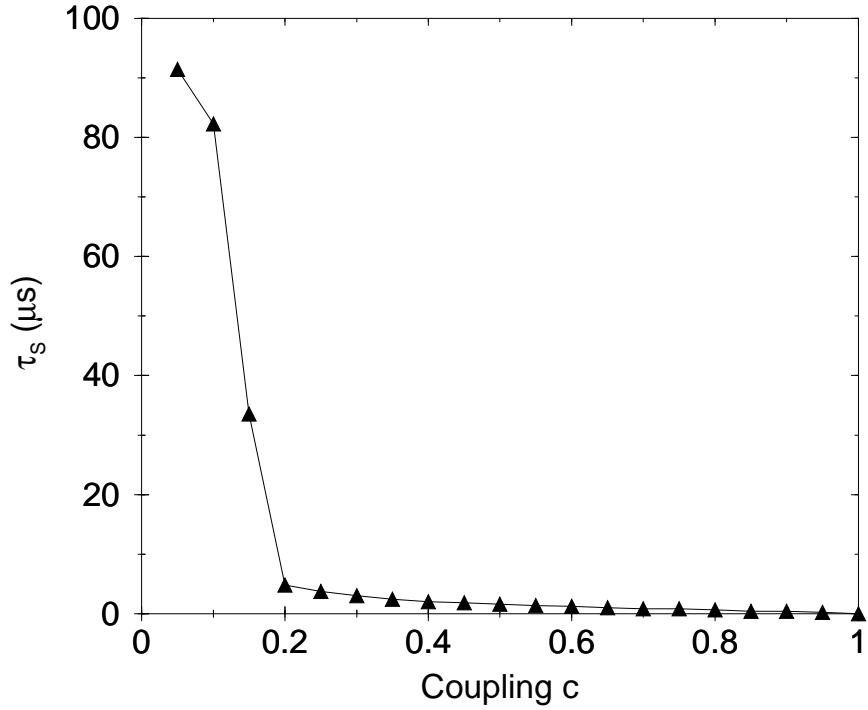,width=4.5in,angle=270}}
\caption{Time to synchronization for identical DFRLs. The synchronization 
time $\tau_s$ is the time at which the amplitude synchronization error 
quantity $H_E(c,t)$ goes smaller than $\epsilon = 10^{-2}$ and stays below 
for all larger times $t > \tau_s$. The value of $\tau_s$ is then averaged 
over 25 different initial conditions.}
\label{fig:amptime_ident}
\end{figure}

The time to synchronization $\tau_s$ is selected so at $t = \tau_s$
the synchronization error $H_E(c,t)$ becomes less than some small
value $\epsilon$, and then stays smaller than $\epsilon$ for all times
$t > \tau_s$. Choosing $\epsilon = 10^{-2}$, we plot $\tau_s$ in
Figure \ref{fig:amptime_ident} as a function of $c$ for $0 \le c \le
1$. We see that for some values of $c \le 0.1$, the lasers do
synchronize, but it takes nearly $100 \mu $s to come within
$\epsilon$. This is much longer than when $c \ge 0.2$, however, where
synchronization sets in rapidly ($\tau_s \le 5 \mu $s). This figure
represents the average time to synchronization over 25 different
initial conditions for each system, thus substantially reducing the effect
of individual trajectory behavior. One lesson we can
draw from Figure \ref{fig:amptime_ident} is that at $c=1$, which is
open loop operation of the receiver, synchronization within $\epsilon$
sets in essentially instantaneously.

This $\tau_s$ is the time at which the synchronization error
$H_E(c,t)$ falls below an arbitrary specified level. This measure
reveals little about the synchronization dynamics beyond this time.
We now examine the residual synchronization error $H_E(c,t)$ for large
times after coupling.  We adopt the general unifying definition of
synchronization proposed by Brown and Kocarev~\cite{brokoc}. Their
formalism states that two subsystems are synchronized with respect to
the subsystem properties $g(x)$ and $g(y)$, if there is a
time-independent function $h$, such that
\be
\label{eqn:norm}
||h[g(x),g(y)]||=0,
\ee
where $||\bullet||$ is some norm. The quantities $g(x)$ and $g(y)$ are
completely general and can refer to any measurable property of the
subsystem. The general form of the time-independent function $h$ which
we adopt here is
\be
\label{eqn:synchdef}
h[g(x),g(y)]\equiv\mathop{\lim}_{T\rightarrow\infty} \frac{1}{T}\int^{t+T}_t\left|g(x(s))-g(y(s))\right|ds.
\ee
For practical reasons, we must modify this statement somewhat, since
numerically we can neither take $T\rightarrow\infty$, nor hope for
$||h[g(x),g(y)]||$ to equal precisely zero, since we will reach the
numerical limits of computation before that occurs.

To keep in the spirit of this definition, we let the coupled chaotic
lasers run for K round trips after coupling, then calculate the RMS
value of $H_E(c,t)$ over M additional round trips to examine the
magnitude of the synchronization error for large times after
coupling. This leads us to the synchronization error function
\be
\label{eqn:identsynch}
D_E(c) = \frac{1}{\Nc_E}\left(\int_{K\tau_R}^{(K+M)\tau_R}H_E^2(c,t)dt\right)^{1/2}.
\ee
The normalization factor $\Nc_E$ is the RMS average for the $c=0$ or uncoupled case 
\be
\Nc_E = \left(\int_{K\tau_R}^{(K+M)\tau_R}H_E^2(c=0,t)dt\right)^{1/2}
\ee
and is included to give us a tangible measure of the magnitude of the
synchronization error at $c$ versus the case of no coupling,
$c=0$. This function compares directly the difference of fields in the
coupled lasers, so for us $g(x)$ and $g(y)$ in equation
(\ref{eqn:synchdef}) are
\bea
g(x(t)) & = & \frac{\Ec_T(t)}{\avabs{\Ec_T}_{\tau_R}} \\
g(y(t)) & = & \frac{\Ec_R(c,t)}{\avabs{\Ec_T}_{\tau_R}}.
\eea
In the work we report here we used $K = 20,000$ and $M = 3,000$.

\begin{figure}
\centerline{\psfig{file=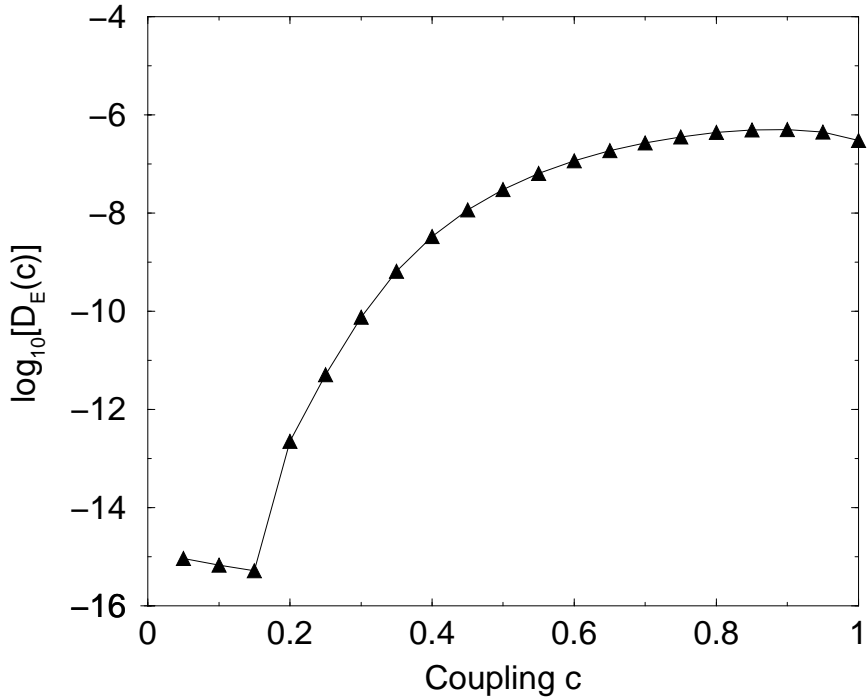,width=4.5in,angle=270}}
\caption{Amplitude synchronization error $D_E(c)$ for identical 
DFRLs. The DFRLs are first coupled for $20,000\tau_R$ and then 
the error term is averaged over an additional $3,000\tau_R$. 
The result is then averaged over 25 initial conditions.}
\label{fig:amprms_ident}
\end{figure}

Fig. \ref{fig:amprms_ident} displays the synchronization error
function $\log_{10}[D_E(c)]$ versus $c$. We see a quite different
picture of the synchronization here.  Above we noted that for coupling
as low as $c=0.1$, the lasers took a much longer time ($\approx 0.1$
ms) than for higher coupling for $H_E(c,t)$ to become as small as
$\epsilon$. However, on the time scale of a few milliseconds, we see
that with much smaller coupling the synchronization error $D_E(c)$ is
less by many orders of magnitude. Note, however, that the error never
exceeds a few parts in $10^7$ for any $c$. Once the lasers synchronize
at some $c$, they do so very accurately.

\begin{figure}
\centerline{\psfig{file=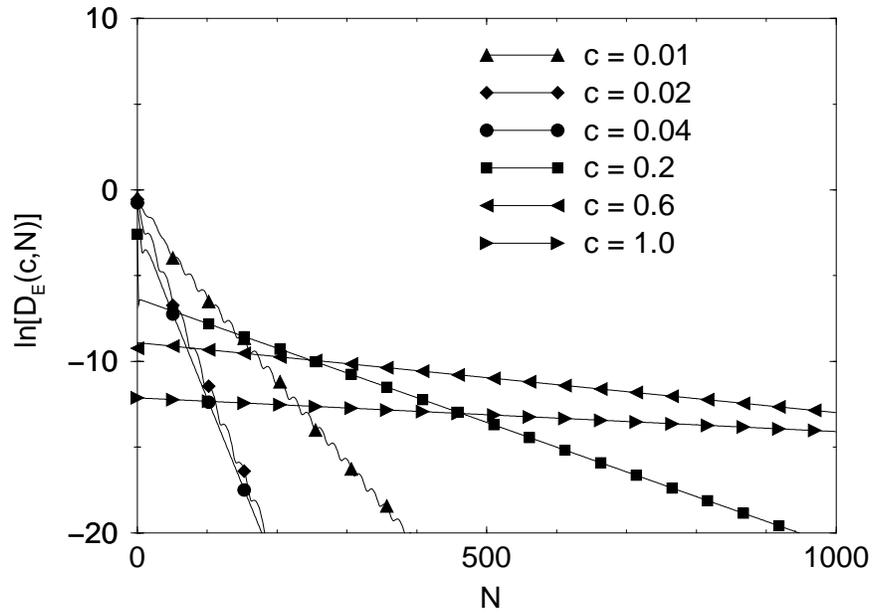,width=4.5in,angle=270}}
\caption{Plot of $D_E(c)$ versus the number of round trips $N$ since
coupling was initiated for coupling constants in the
range of $0.01 \le c \le 1.0$. The lasers are identical.}
\label{fig:syncerror}
\end{figure}

To investigate these results quantitatively, we look at a plot which
shows the temporal behavior of the synchronization error
$D_E(c)$. We reintroduce a discrete time variable
$N\tau_R$, where $N$ is the number of round trips since coupling was
initiated, and $\tau_R$ is the round trip time. We then define
$D_E(c,N)$ as the synchronization error averaged over a round trip
\be
D_E(c,N) = \frac{1}{\Nc_E}\left(\int_{(N-1)\tau_R}^{N\tau_R}H_E^2(c,t)dt\right)^{1/2}.
\ee
Fig. \ref{fig:syncerror} shows the evolution of 
$\ln(D_E(c,N))$ versus $N$ for couplings in the
range $0.01 \le c \le 1.0$. We see a
pattern in convergence rates with varying coupling (the
rate of convergence is the slope of the $\ln(D_E)-N$
plot). After an initial rapid jump of $D_E(c,N)$ towards
synchronization in the first few round trips, the weaker coupling
cases converge at an increasingly faster rate (they have steeper
negative slopes), with the slowest convergence rate occurring at the
strongest coupling, $c=1$. 

There are thus two time scales of
synchronization. There is an initial time scale ($t < 1\mu
$s) with a rapid, short term convergence toward
synchronization, then slower convergence rate 
for longer times ($t>1\mu$s). For weaker coupling,  the
initial convergence of the lasers is not as large as
for the stronger coupling case, since the
amount of optical field from the transmitter being introduced into the
receiver is proportional to the coupling strength. However, past this
initial short time scale, the weaker coupling draws together
the two lasers' trajectories at a faster rate than for the stronger
coupling.
 
\begin{figure}
\centerline{\psfig{file=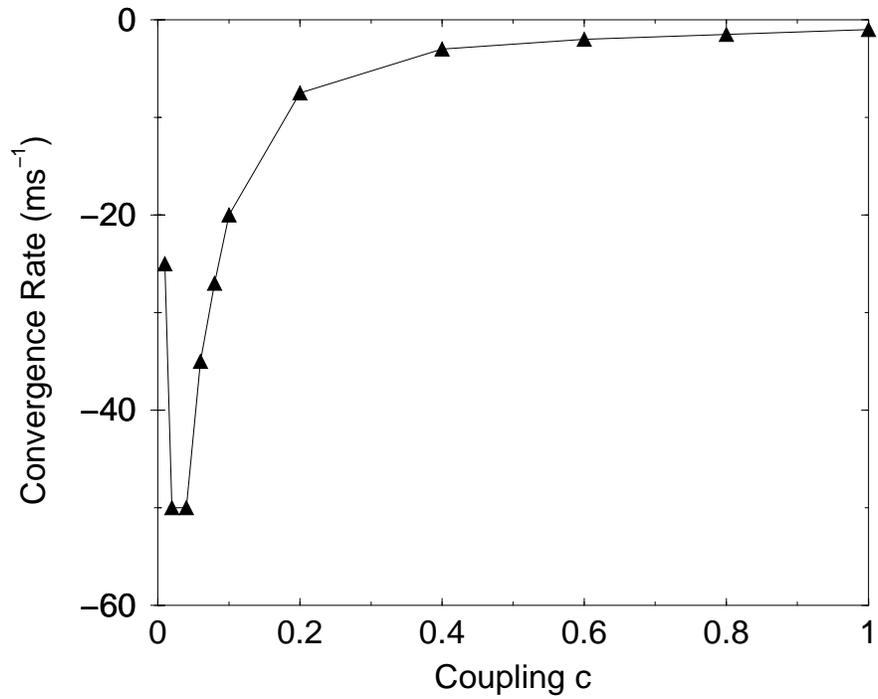,width=4.5in,angle=270}}
\caption{Plot of the rate of convergence for the measure $D_E(c)$ for
a range of coupling constants. The slope is calculated beginning after 
the initial convergence upon coupling. The lasers are identical.}
\label{fig:syncslopes}
\end{figure}

Examining the weak coupling range, $0.01 \le c \le 0.04$  in
Fig.~\ref{fig:syncerror}, we see that the magnitude of the
convergence rate is maximal between $0.02 \le c \le 0.04$.  Figure
\ref{fig:syncslopes} shows the rate of convergence plotted against
coupling constant c.  The coupling is linear in $\Ec_T(t)$ and
$\Ec_R(t)$, so this complex synchronization behavior is not likely due
to this mixing of the optical fields alone, therefore we look for
explanations outside of the optical field variables. The other system
variables are the population inversions $w_T(t)$ and $w_R(t)$. They
cannot be directly coupled, being related only indirectly through the nonlinear
relationship to their respective internal optical intensities
(equation (\ref{eqn:popinv})).  

\begin{figure}
\centerline{\psfig{file=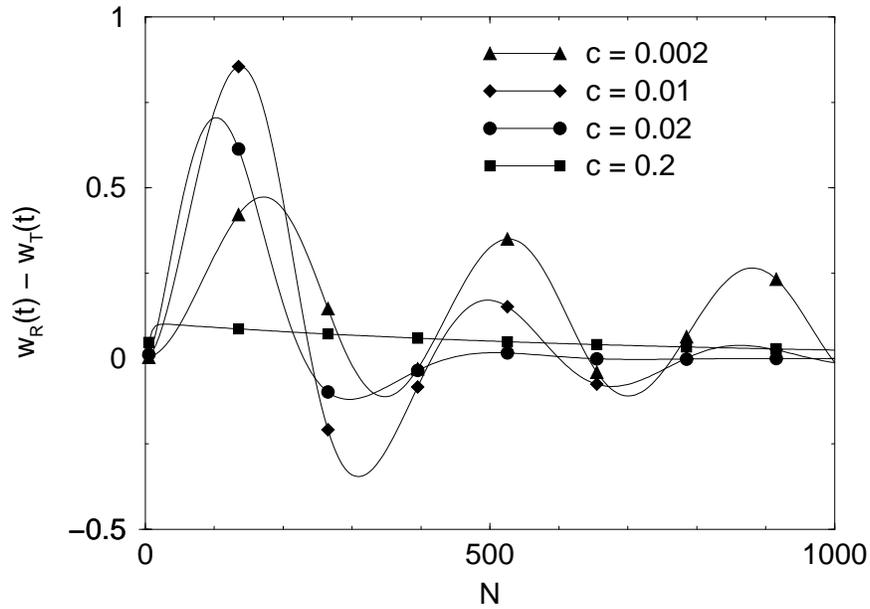,width=4.5in,angle=270}}
\caption{The dynamics of the integrated population inversion plotted against 
the number of round trips $N$ since coupling was initiated in the range 
$0.002 \le c \le 0.2$. The lasers are identical.}
\label{fig:w_dynamics}
\end{figure}

The time dependence of $w_R(t)$ and $w_T(t)$ reveals insight into the
dynamics just after the lasers are coupled. In
Fig.~\ref{fig:w_dynamics} we show $w_R(t)-w_T(t)$ as a function of
time after the lasers are coupled and as a function of $c$. In the
transmitter, $w_T(t)$, is essentially constant because erbium's
fluorescence lifetime $T_1$ is so long ($T_1 \approx\,$ 10 ms) so that
in the long-term asymptotic state of the laser, changes in $w_T(t)$ occur only on the order
of $10^5$ round trips.  There is always a rise in $w_R(t)$
immediately after coupling because when $0 <  c < 1$, the intensity
in the receiver laser just as the lasers are coupled is $|c\Ec_T|^2 +
|(1-c)\Ec_R|^2 + 2c(1-c)\Re(\Ec_T\cdot\Ec_R^*)$ averaged over a round
trip, and the cross term averages to zero since the fields are
initially uncorrelated. As $\avabssq{\Ec_T}$ and $\avabssq{\Ec_R}$
averaged over a round trip are equal, the average
intensity initially entering the receiver's active medium will be less
that $\avabssq{\Ec_R}$ by a factor of $c^2+(1-c)^2;\; 1/2 \le
c^2+(1-c)^2 \le 1$. This allows the pumping term to increase
the receiver population inversion as the active medium sees a reduced
intensity trying to stimulate transitions between lasing states.

This effect continues only as long as $\Re(\Ec_T\cdot\Ec_R^*) \ll
|\Ec_T||\Ec_R|$.  For large coupling, correlation of the two fields
occurs within a few round trips, before the population inversion has a
chance to increase much.  However, for small coupling, it takes
several hundred round trips for the two lasers' fields to start
becoming correlated, so the population inversion has the chance to
grow substantially. This in turn will cause the average electric field
intensity in the receiver laser to grow.  

We see in Fig. \ref{fig:w_dynamics} for $c=0.002$, that $w_R (t)$
oscillates in a manner reminiscent of an under-damped oscillator.
This coupling coefficient is an order of magnitude smaller than the region
we found to contain the fastest convergence rate and the figure
verifies that the oscillations are slow to decay, thereby causing the
approach to synchronization to also be slow. In the same figure for
$c=0.01$, which is closer to the optimal coupling region, and we can
see that the oscillations are still present, but are decaying much
faster. As we reach the optimal convergence region at $c=0.02$, we see
that the decay now takes only a little more than an oscillatory cycle
to damp considerably.  This would correspond to the near critically-damped
case.  Then as we increase coupling further to $c=0.2$, the
oscillations are gone but now after the initial rise in $w_R(t)$, it
again takes much longer to decay, much like an over-damped oscillator.

To finish the examination of the case of identical subsystems, we next
determine the minimal value of $c$ which leads to synchronization. In
the discussion above the lasers always synchronize, so now we examine
weaker coupling yet. To do this we numerically evaluate the largest
conditional Lyapunov exponent~\cite{pc1} in much the same way as we
find the standard largest Lyapunov exponents in our earlier
paper~\cite{akbl1}.  Now we couple the identical lasers with small
coupling $c$ and ask when the largest conditional exponent becomes
negative as we vary $c$. The critical value of coupling was found to
be $c_{crit} \approx 1.3 \times 10^{-3}$. Therefore, for $c <
c_{crit}$, we observe no synchronization, as a positive conditional exponent
implies loss of synchronization. 

\subsection{Synchronization at c = 1}

We analytically examine our coupled transmitter and receiver system at
$c=1$, that is, when we run the receiver open loop. This is the
configuration in the Georgia Tech
experiments~\cite{science,prlrvw,private}. In this case the field injected
into the receiver is just $\Ec_T(t)$ and none of the field in the
receiver fiber is re-injected into the amplifier. The equations for
the two coupled systems are
\begin{itemize}
\item For the Transmitter Laser
\bea
\Ec_T(\tau + \tau_R) &=& \M_{\Ec}(w_T(\tau),\Ec_T(\tau)) \nonumber \\
\frac{d w_T(\tau)}{d \tau} &=& 
Q - \gamma \biggl \{w_T(\tau) + 1 + |\Ec_T(\tau)|^2(e^{G
w_T(\tau)} -1) \biggr \} 
\eea
\item For the Receiver Laser

\bea
\Ec_R(\tau + \tau_R) &=& \M_{\Ec}(w_R(\tau),\Ec_T(\tau)) \nonumber \\
\frac{d w_R(\tau)}{d \tau} &=& 
Q - \gamma \biggl \{w_R(\tau) + 1 + |\Ec_T(\tau)|^2(e^{G
w_R(\tau)} -1) \biggr \}, 
\eea

\end{itemize}
where $\M_{\Ec}(w(\tau),\Ec(\tau))$ is the map defined in earlier
sections.  Note that in the receiver equations only $\Ec_T(\tau)$ now
appears on the right hand side.
Take the difference of the population inversion equations to arrive at
\be
\frac{d(w_T(\tau)-w_R(\tau))}{d \tau} =
-\gamma\biggl\{w_T(\tau)-w_R(\tau)
+|\Ec_T(\tau)|^2e^{Gw_R(\tau)} (e^{G(w_T(\tau)-w_R(\tau))}
-1)\biggr\}.
\ee
Noting that $e^x-1 \ge x$, we can write the following inequality
\be
\frac{d(w_T(\tau)-w_R(\tau))}{d \tau} \le -\gamma\left(w_T(\tau)-w_R(\tau)
\right) \biggl\{1 +|\Ec_T(\tau)|^2e^{Gw_R(\tau)} \biggr\}.
\ee
This shows that $w_T(\tau)-w_R(\tau)$ goes to
zero exponentially rapidly at a rate governed by $\gamma (1
+|\Ec_T(\tau)|^2e^{Gw_R(\tau)}) $.  This result on the synchronization
at $c=1$ is a {\bf global} property of these laser systems. Nowhere
was a linearization made about the synchronization manifold.

This value of the convergence rate to synchronization agrees within
$0.1\%$ of the numerical calculation of the same convergence rate of
$w_T(\tau)-w_R(\tau)$ at $c=1$ in our numerical simulations. This gives us
additional confidence in both the simulations and in the details of
the approximations which went into evaluating the propagation of light
around the fiber ring with nonlinear effects~\cite{akbl1}.

The final step is to use this bounded behavior of the difference in
population inversions in the maps
\bea
\Ec_T(\tau + \tau_R) &=& \M_{\Ec}(w_T(\tau),\Ec_T(\tau)) \nonumber \\
\Ec_R(\tau + \tau_R) &=& \M_{\Ec}(w_R(\tau),\Ec_T(\tau)).
\eea
With this one easily shows that as $w_T(\tau)-w_R(\tau) \to 0$ so does
$\Ec_T(\tau) - \Ec_R(\tau) \to 0.$ This result demonstrates {\em
global} stability of the synchronization manifold $w_T(\tau) =
w_R(\tau)$ as it involves no linearization of the equations around
this solution. It is the detailed structure of the DFRL equations
which permits this demonstration of global stability of the
synchronization manifold.

The strong rate of convergence of $w_T(\tau)-w_R(\tau)$, approximately
as $\exp(-\gamma|\Ec_T|^2\tau),$ implies that small perturbations
to synchronization which might arise due to noise in the channel or
disturbances of the receiver would rapidly be `cured' by the
auto-synchronization nature of the system at $c=1$. This attractive
robustness also suggests that small mismatches in parameters of the
transmitter and receiver will also affect the synchronization only
slightly. In the next section we show that we indeed have this feature
of robust synchronization when we have mismatches in many of the
system parameters. An important exception occurs when we have a
mismatch in any of the parameters which deal with the complex
vectorial nature of the optical field.

\subsection{Mismatched Transmitter and Receiver}

When the transmitter and receiver lasers are identical with no noise
or other disturbance in the transmission of $\Ec_T(t)$ from
transmitter to receiver, identical synchronization occurs rapidly for
extremely small values of $c$. 

We now examine the influence of system mismatch on synchronization
performance.  Our numerical investigation proceeds much as before. We
run each laser for 400,000 $\tau_R$, then we couple the two lasers
with some value of $c;\;0\le c\le1$. The lasers are permitted to
attempt to synchronize for 20,000 $\tau_R$ and then the statistics are
evaluated over the next 3,000 $\tau_R$.

With parameter mismatch, identical synchronization is no longer
expected because the transmitter subsystem is no longer identical to
the receiver subsystem, and thus identically synchronized motion is
not a solution of the receiver dynamics.  This is not
necessarily devastating news, however. The error measure $D_E(c)$
quantifies the identical synchronization of the lasers, i.e., the
measurable variables $g(x)$ and $g(y)$ in equation
(\ref{eqn:synchdef}) are the entire two-dimensional, complex vector
values of the optical field.  Two identically synchronized lasers have
synchronous amplitude, polarization, and phase fluctuations. However,
to communicate via synchronization, we need not satisfy such rigid
requirements.

For example, the main method of encoding a message onto the output of
the transmitter laser in the work of VanWiggeren and
Roy~\cite{science} has been Amplitude Shift Keying (ASK). The
intensity of the transmitter laser is electro-optically modulated
according to the desired bit pattern.  Dividing the incoming optical
field intensity by the intensity of the synchronized receiver laser
results in the recovery of the original bit sequence. In this
experiment the lasers are run in open loop operation ($c=1$) and thus
parameter mismatches in the receiver and transmitter have minimal
effect.  Parameter mismatch for $c < 1$ almost invariably alters the
time-asymptotic mean intensity of the lasers because this mean
intensity is a parameter-dependent quantity. This means that the field
intensity relationship between the two lasers can not remain
$|\Ec_T(t)|^2=|\Ec_R(c,t)|^2$.  However, if the field intensity
relationship between the two lasers converges to a form
$\alpha(c)|\Ec_T(t)|^2=|\Ec_R(t)|^2$, where $\alpha(c)$ is a constant
for each $c$, then we could say that the intensities of the two laser
subsystems are in a state of intensity synchronization. In this case,
with knowledge of $\alpha(c)$, the ASK decoding could still be
performed and suitable message recovery achieved.

For this reason we turn now to an examination of the possibility that 
perhaps only some of the measurable properties of the subsystems are 
synchronized. This phenomenon falls in the category of what has been 
termed {\em generalized synchronization}~\cite{rsta}.

\subsubsection{Generalized Synchronization}

To make the following discussion compatible with the previous
discussion of synchronization, we cast generalized synchronization in
the language of equation (\ref{eqn:norm}). Two subsystems of coupled
dynamical systems are considered to be in generalized synchronization
if there is a comparison function $h$ given by
\be
h[g(x),g(y)] = \mathop{\lim}_{T\rightarrow\infty}\int_t^{t+T}\left|H[g(x)]-g(y)\right|ds
\ee
that satisfies equation (\ref{eqn:norm}). $H(\bullet)$ is a
smooth, invertible, time-independent
function~\cite{kocpar,pjk,hoy,brown}. This would imply that if $g(y(t))
= H[g(x(t))]$ as $t\rightarrow\infty$, then we have generalized
synchronization. Examples have been found where generalized
synchronization exists, but where $H$ is not an invertible operation
\cite{rsta,rulsush}.

For our purposes we do not need to delve too deeply into the nuances
of this formalism, because what we want to do is much simpler. All that we
require is a method to compare the values of different physical,
experimentally measurable properties between the transmitter and
receiver DFRLs. Therefore we can still use our general definition of
synchronization from equation (\ref{eqn:synchdef}) with a
generalization of the property comparison function on the inside of
the integral.  This form is
\be
h[g(x),g(y)]\equiv\mathop{\lim}_{T\rightarrow\infty} \frac{1}{T}\int^{t+T}_t H[g(x),g(y)] ds.
\ee
where $H[g(x),g(y)]$ is now a general function comparing the
measurable subsystem properties $g(x)$ and $g(y)$, and it depends upon
the specific property that is meant to be compared.

To look for generalized intensity synchronization, we define a
comparison function:
\be
H_I[|\Ec_T(t)|^2,|\Ec_R(c,t)|^2] = \log_{10}\left(\frac{|\Ec_R(c,t)|^2}{|\Ec_T(t)|^2}\right).
\ee
The logarithm here is included to remove the bias caused by
taking the ratio of the intensities. For example, if the intensities
were in perfect generalized synchronization, i.e.,
$\alpha(c)|\Ec_T(c,t)|^2=|\Ec_R(t)|^2$, the plot of
$|\Ec_T(t)|^2$ versus $|\Ec_R(c,t)|^2$ would be a perfect straight
line with slope $\alpha(c)$. However, if the synchronization were not
perfect, there would be data points off of the line with slope
$\alpha(c)$. If enough points were off the line, then it would no
longer be clear what was the correct value for $\alpha(c)$. The
natural procedure then would be to take an average over the data set
to find $\alpha(c)$. The problem arises because each data point in the
average represents a ratio of receiver intensity to transmitter
intensity. An arithmetic average of the ratio data points will bias the
average toward the larger ratios. For example, a ratio of
\be
\frac{|\Ec_R(c,t)|^2}{|\Ec_T(t)|^2}=\frac{2}{1}
\ee
will effect the average of $\alpha(c)$ upward more than a ratio of 
\be
\frac{|\Ec_R(c,t)|^2}{|\Ec_T(t)|^2}=\frac{1}{2}
\ee
will effect it downwards. However, in the two-dimensional phase space
of $|\Ec_T|^2$ and $|\Ec_R|^2$, the points $(2,1)$ and $(1,2)$ are the
same distance from the $45^{\circ}$ line. Therefore, we need a
comparison function which removes this bias and treats the averages
equally in their phase space distance interpretation. The logarithm
function does this automatically since $\log_{10}(\alpha) =
-\log_{10}(1/\alpha)$.

Using this comparison function, we introduce a generalized
synchronization error measure $D_I(c)$ analogous to $D_E(c)$ in
equation (\ref{eqn:identsynch}). However, we highlight one important
difference. In equation (\ref{eqn:identsynch}) we were looking for a
trend toward identical synchronization, so we took an RMS error
average to calculate the variance around $H_E(c,t)=0$. In the
generalized synchronization case, we do not expect $H_I(c,t)=0$, but
convergence to some asymptotic value of $\alpha(c)$. Thus 
we measure the standard deviation of $H_I(c,t)$
about $\alpha(c)=\av{H_I(c,t)}_{M\tau_R}$:
\be
D_I(c) = \frac{1}{\Nc_I}\left(\int_{K\tau_R}^{(K+M)\tau_R}\left[H_I(c,t) - 
\av{H_I(c,t)}\rangle_{M\tau_R}\right]^2 dt\right)^{1/2}.
\ee
where
\be
H_I(c,t) = \log_{10}\left(\frac{|\Ec_R(c,t)|^2}{|\Ec_T(t)|^2}\right)
\ee
Again this integral is calculated for $K=20,000$ and $M=3,000$ and 
normalized by the factor $\Nc_I = D_I(c=0)$. 

There are many other generalized synchronization relationships which 
could be exploited for specific communication methods. Another possibility 
would be encoding a message with polarization modulation~\cite{betti}. 
If the polarization evolution is synchronized, comparison of the two lasers' 
state of polarization could result in successful recovery of the message. 
To examine the possibility of generalized polarization synchronization, we 
introduce a new comparison function
\be
H_{\theta}(c,t) = \theta_S (c,t),
\ee
where $\theta_S(c,t)$ is the angle in Stokes parameter space between the two states 
of polarization in the two laser subsystems. This leads to a
synchronization error measure analogous to $D_I(c)$, calculating now the 
time average of $\theta_S(c,t)$,
\be
D_{\theta}(c) = \frac{1}{\Nc_{\theta}}\left[\int_{K\tau_R}^{(K+M)\tau_R}\left(H_{\theta}(c,t)-\av{H_{\theta}(c,t)}_{M\tau_R}\right)^2 dt\right]^{1/2}.
\ee
The normalization is again $\Nc_{\theta} = D_{\theta}(c=0)$. $\theta_S$
is found using the Stokes parameters~\cite{betti}
\bea
S_0 & = & a^2+b^2+c^2+d^2 \\
S_1 & = & a^2+b^2-c^2-d^2 \\
S_2 & = & 2(ac+bd) \\
S_3 & = & 2(ad-bc),
\eea
for an electric field with arbitrary x and y polarization
\be
\Ec = (a+ib)\hat{x} + (c+id)\hat{y}.
\ee

These satisfy
\be
S_0^2 = S_1^2 + S_2^2 + S_3^2,
\ee
so the state of polarization can be represented as a vector $\vec{S}$ 
in ($S_1,S_2,S_3$) space of magnitude $S_0$. The angle between the two 
states of polarization of $\Ec_T$ and $\Ec_R$ is then
\be
\theta_S = \cos^{-1}\left[\frac{\vec{S_T} \cdot \vec{S_R}}{S_{0_T}S_{0_R}}\right].
\ee 
This is the value which is time averaged to examine if the lasers are in
a state of generalized polarization synchronization.

Finally, we examine another class of generalized synchronization
potentially useful for communications: optical phase
synchronization. If the phase of the two lasers is synchronized,
communication through Phase Shift Keying (PSK)~\cite{betti} would be
possible. To look for phase synchronization, we introduce the
comparison function,
\be
H_{\phi}(c,t) = \phi_T - \phi_R,
\ee
where $\phi_T$ and $\phi_R$ are defined in detail below. This leads to another error measure
\be
D_{\phi}(c) = \frac{1}{\Nc_{\phi}}\left[\int_{K\tau_R}^{(K+M)\tau_R}\left(H_{\phi}(c,t)-\av{H_{\phi}(c,t)}_{M\tau_R}\right)^2 dt\right]^{1/2}.
\ee
Again, $\Nc_{\phi} = D_{\phi}(c=0)$. 

Each electric field $\Ec_T$ and $\Ec_R$ is in general elliptically
polarized. We desire to quantify the phase with respect to its own
polarization basis, so we define the phase of the electric field as
its phase in the x-y laboratory frame minus the angle of the major
axis of the polarization ellipse with respect to the x-y laboratory
frame, i.e.,
\be
\phi = \phi_{xy} - \Phi_{ellipse} \\\ (\mbox{mod}\, \pi).
\ee
Where, using the above definitions, we have
\be
\phi_{xy} = \tan^{-1}\left[\frac{c}{a}\right]
\ee
and
\be
\Phi_{ellipse} = \frac{1}{2}\tan^{-1}\left[\frac{S_2}{S_1}\right].
\ee

\begin{figure}
\centerline{\psfig{file=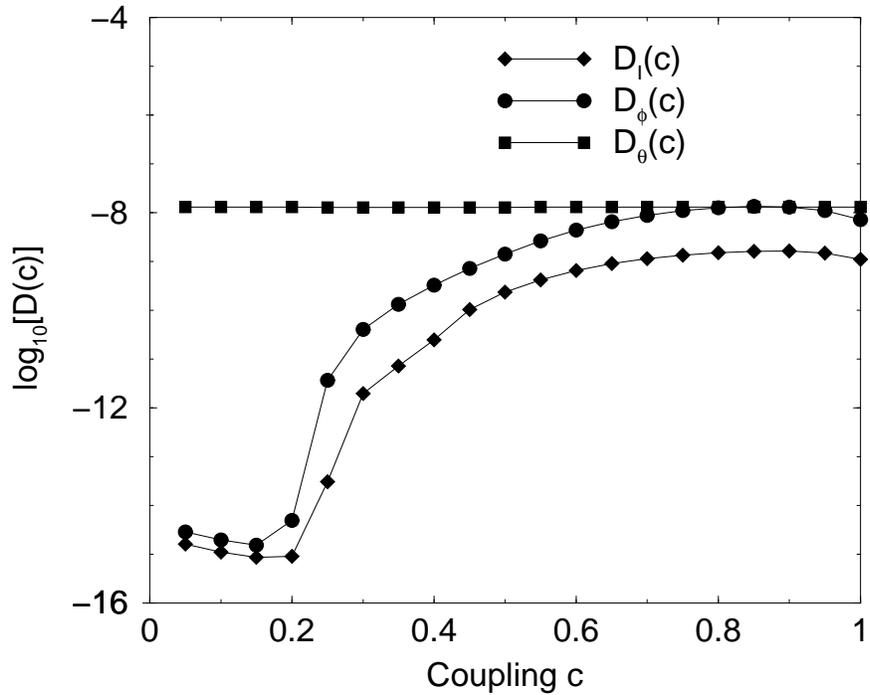,width=4.5in,angle=270}}
\caption{Generalized synchronization measures $D_I(c)$, $D_{\phi}(c)$,
and $D_{\theta}(c)$ plotted against coupling constant c. The
DFRLs are coupled for $20,000\tau_R$ and the measures are taken over
an additional $3,000\tau_R$. The lasers are identical.}
\label{fig:ident_general}
\end{figure}

Armed with this array of error measures for the various classes of
generalized synchronization we now move on to examine the effects of
parameter mismatch. We provide an idea of the scale of these error
measures by displaying their values for the case of {\em identical
lasers}. In Fig.~\ref{fig:ident_general} we show $D_I(c)$,
$D_{\phi}(c)$, and $D_{\theta}(c)$. Since identical lasers exhibited
strong identical synchronization, the generalized synchronization
errors are extremely small as well. For example, the standard
deviation of the generalized intensity $D_I(c)$ starts at $10^{-15}$
for $c=0.05$ and increases to about $10^{-9}$ at $c=1$. The lower
coupling ranges push the limits of our computational accuracy, so
although the quantitative values can not be trusted, they indicate
that the error is effectively zero for couplings $c \le 0.2$. The
standard deviation of the phase $D_{\phi}(c)$ is similarly small
ranging from $10^{-15}$ to $10^{-8}$. Again this tests the limits of
numerical accuracy for small couplings. The polarization deviation
$D_{\theta}(c)$ is reasonably constant at a very low error of
$10^{-8}$.

\subsubsection{Gain Mismatch}

First we introduce mismatch in the gain term $G$ 
of the rare earth amplifiers. We examined synchronization as a function of the dimensionless ratio 
\be
{\cal{G}} = \frac{|G_R - G_T|}{G_T}
\ee
over the range 10\% to 50\%. Mismatches in the gain could arise from a
difference in the length of rare earth doped fiber in the amplifiers or a
difference in doping level or perhaps a difference in absorption of light at
1.55 $\mu m$ in the particular doped fibers used in the amplifiers.

\begin{figure}
\centerline{\psfig{file=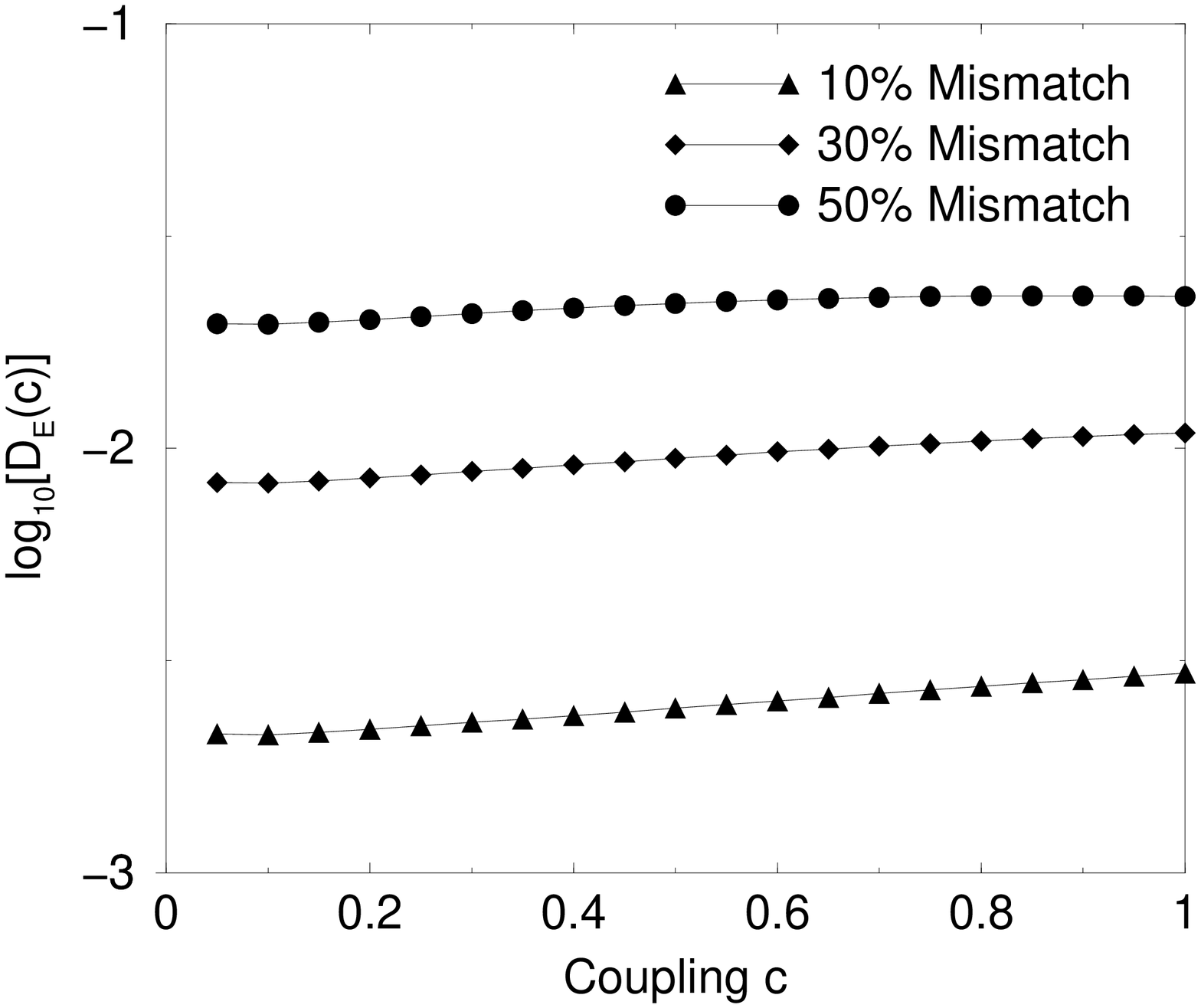,width=4.5in,angle=270}}
\caption{Identical synchronization error $D_E(c)$ for DFRLs with gain 
mismatch $\cal{G}$. The DFRLs are first coupled for $20,000\tau_R$ and then 
the error term is calculated over an additional $3,000\tau_R$.}
\label{fig:gain_amprms}
\end{figure}

\begin{figure}
\centerline{\psfig{file=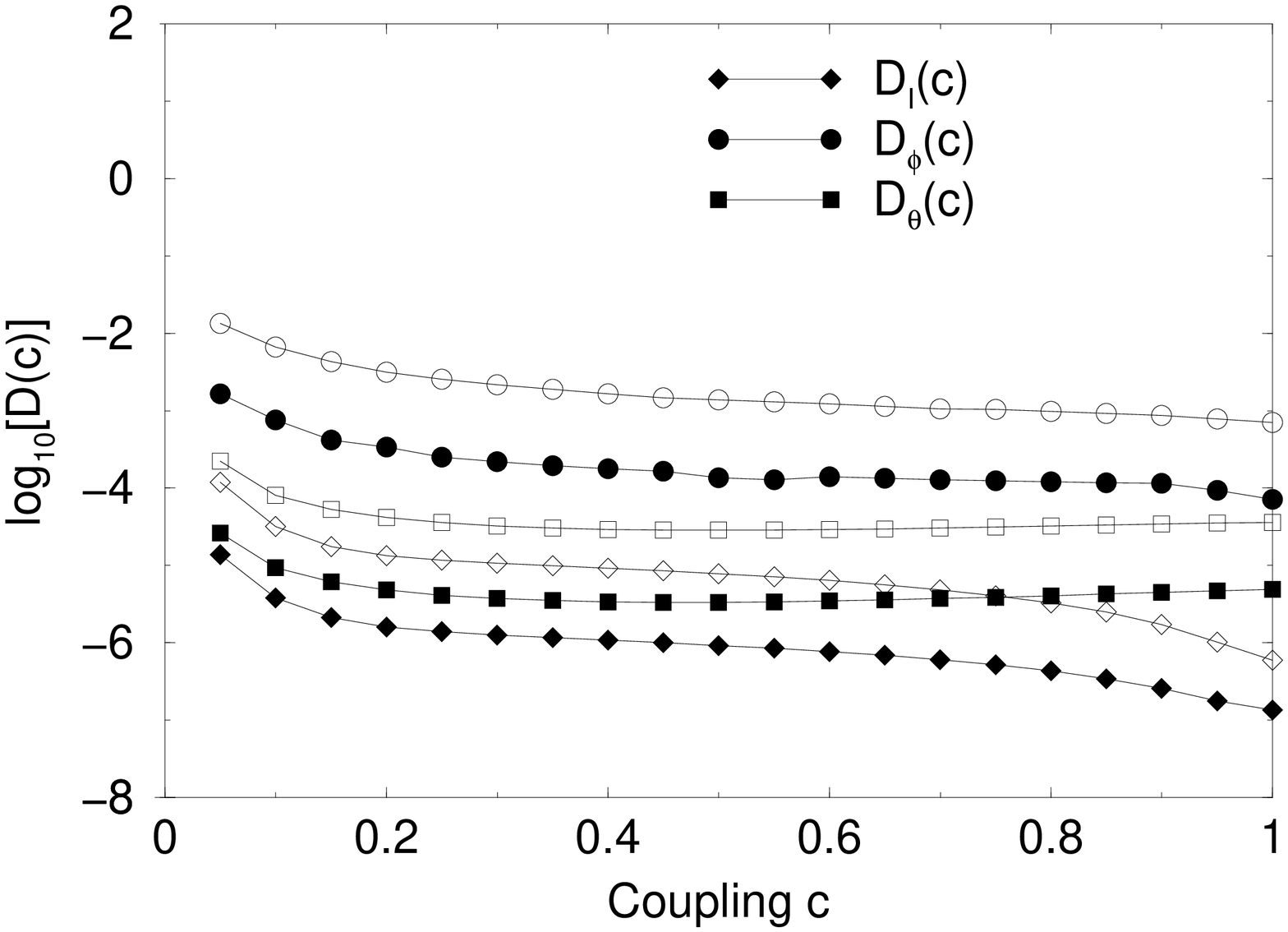,width=4.5in,angle=270}}
\caption{Generalized synchronization measures plotted against coupling constant c. 
There is gain mismatch. The filled symbols represent ${\cal{G}}=10\%$ and the clear
symbols ${\cal{G}}=50\%$. The DFRLs are coupled for 
$20,000\tau_R$ and the measures are taken over an additional $3,000\tau_R$.}
\label{fig:gain_general}
\end{figure}

The identical synchronization measure $D_E(c)$ is shown in 
Fig. \ref{fig:gain_amprms} as a function of $c$ for various $\cal{G}$. Here
the synchronization is only slightly affected by the parameter
mismatch. We see that even for a 50\% mismatch, which is quite large,
the lasers still synchronize within a few parts in $10^2$. We note
that the favorable effect of weaker coupling is still present here in
the presence of mismatch, although much less dominant. Examination of
the generalized synchronization measures $D_I(c)$, $D_{\phi}(c)$, and
$D_{\theta}(c)$ in Fig. \ref{fig:gain_general} reveals that the
quality of synchronization is even more robust for the generalized
cases. Even for ${\cal{G}} = 50\%$ we see intensity synchronization
errors ranging from $10^{-4}$ at weak coupling to $10^{-6}$ close to
$c=1$. The generalized phase and polarization synchronization errors
are both two or more orders of magnitude below unity for the whole
range of $c$. This suggests that in a realistic application,
reasonable mismatches in the gains of the two lasers would be of
minimal concern.

\subsubsection{Pump Mismatch}

\begin{figure}
\centerline{\psfig{file=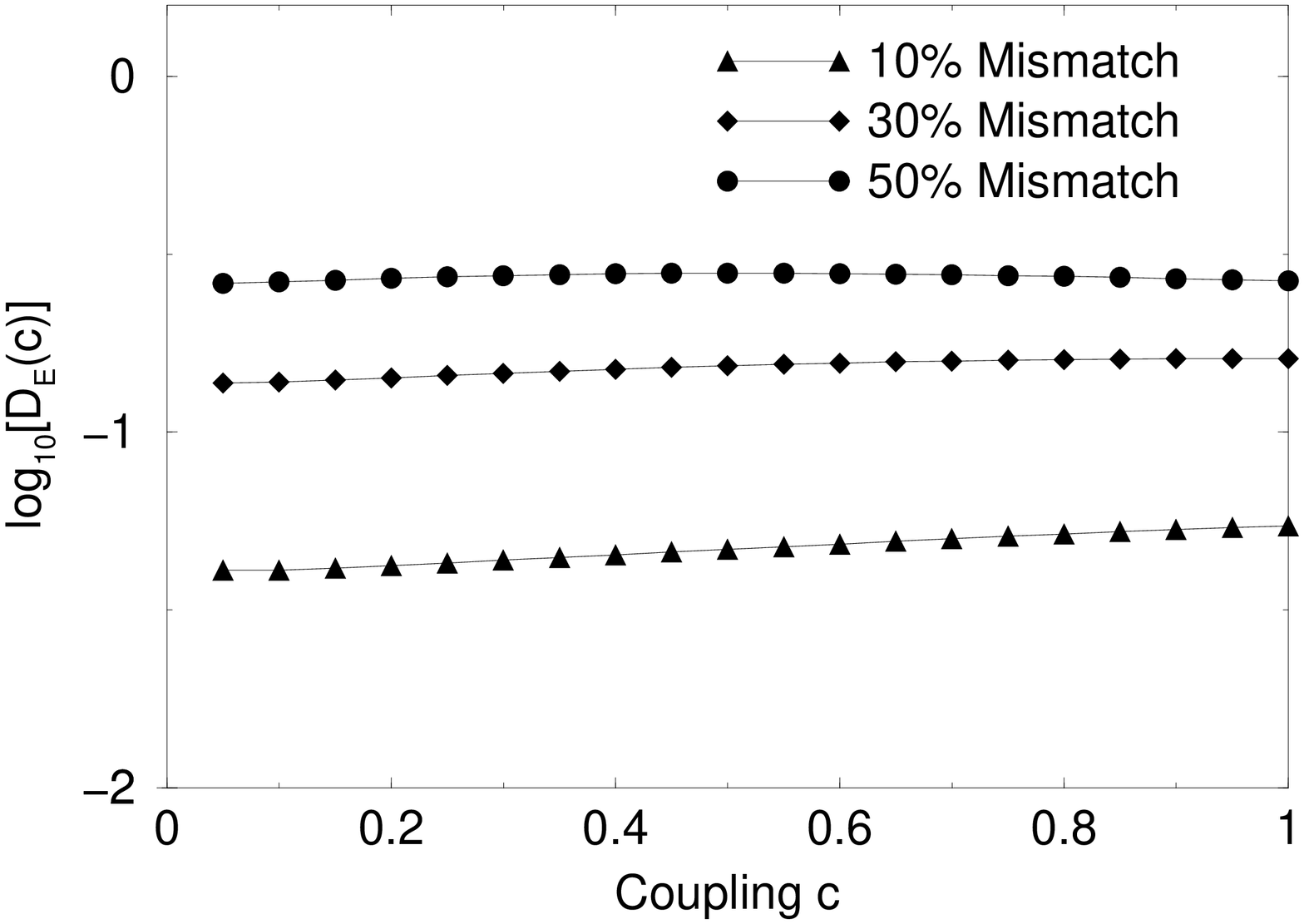,width=4.5in,angle=270}}
\caption{Identical synchronization error $D_E(c)$ for DFRLs with pump 
mismatch $\cal{Q}$. The DFRLs are first coupled for $20,000\tau_R$ and 
then the error term is calculated over an additional $3,000\tau_R$.}
\label{fig:pump_amprms}
\end{figure}

Next we investigated mismatch in the pump levels of the rare earth
amplifiers. The pumping in the experiments of Roy and
VanWiggeren~\cite{science} used a diode laser operating at 980 nm, and
it is unlikely that the pump lasers could be identical in operating
characteristics, so an examination of the effect of pump mismatch has
important physical motivation. In Fig. \ref{fig:pump_amprms} we
display $D_E(c)$ for pump level mismatches of the quantity
\be
{\cal{Q}} = \frac{|Q_R - Q_T|}{Q_T}
\ee
ranging again from 10\% to 50\%. These are quite substantial
mismatches in transmitting and receiving lasers. The effect of pump
mismatch $\cal{Q}$ is reminiscent of that of the gain mismatch, only
more severe. The identical synchronization measure degrades with
increasing mismatch. However for the smaller mismatches, the lasers
still synchronize within a few parts in $10^2$, with the error
becoming much larger than a part in 10 as mismatch is increased to
50\%.

\begin{figure}
\centerline{\psfig{file=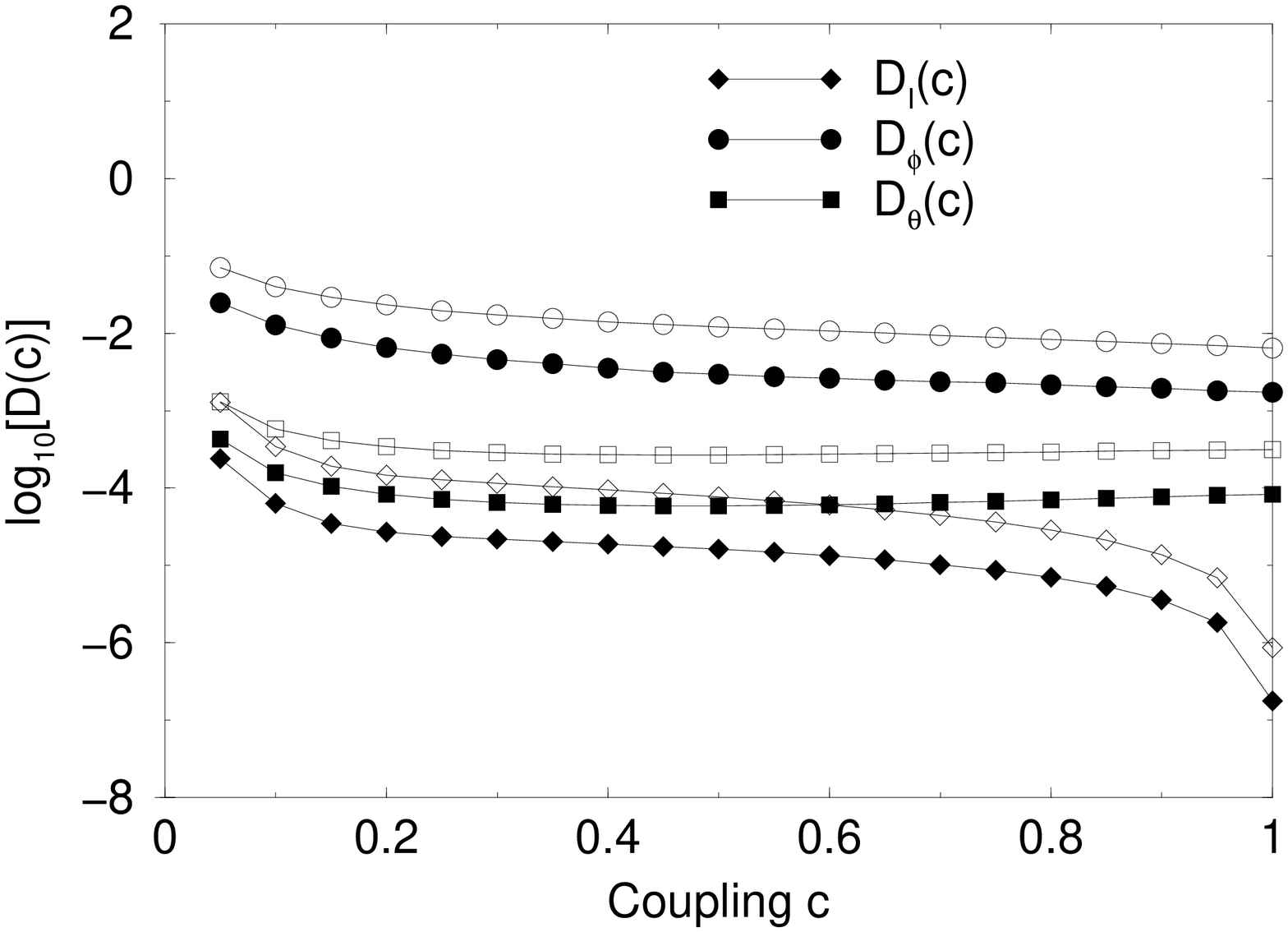,width=4.5in,angle=270}}
\caption{Generalized synchronization measures plotted against coupling constant c. 
There is pump mismatch. The filled symbols represent ${\cal{Q}}=10\%$ and the clear
symbols ${\cal{Q}}=50\%$. The DFRLs are coupled for 
$20,000\tau_R$ and the measures are taken over an additional $3,000\tau_R$.}
\label{fig:pump_general}
\end{figure}

However, the generalized synchronization measures in Fig.
\ref{fig:pump_general} again reveal
very good generalized synchronization. The intensity and the
polarization measures are several orders of magnitude below unity for
the whole range of coupling strengths, although the phase measure is
starting to increase. For the larger mismatch values, the phase
synchronization error is up to a few percent of the uncoupled
normalization value.

Thus, while parameter mismatch adversely affects the quality of
identical synchronization, we see that generalized synchronization
persists, and thereby still provides usable synchronization of
physically measurable variables, in this case the laser
intensities and states of polarization.

\subsubsection{Polarization Controller Mismatch}

Now we examine mismatch in the polarization propagation
characteristics in the transmitter and receiver fibers. As the light
travels down the fiber, its state of polarization varies substantially
over distances of order 10 m. The net effect of all fiber
birefringence, whatever they are, may be represented by a single Jones
matrix. Here, we model the fiber as the overall Jones matrix composed
of a quarter wave plate followed by a half wave plate then followed by
a second quarter wave plate. Up to an overall phase term these can
produce any state of polarization from any other state.  The Jones
matrices for the quarter wave plate is
\begin{equation}
\J_{1/4}(\theta) =
\left(\frac{-i}{\sqrt{2}}\right)\pmatrix{
\cos{(2\theta)} + i & \sin{(2\theta)} \cr
    &     \cr
\sin{(2\theta)} & -\cos{(2\theta)} + i \cr }
\end{equation}
and for the half-wave plate,
\begin{equation} \J_{1/2}(\theta) =
\left(-i\right)\pmatrix{
\cos{(2\theta)} &  \sin{(2\theta)}  \cr
    &    \cr
\sin{(2\theta)} & -\cos{(2\theta)} \cr }.
\end{equation}
In our calculations we use the product Jones matrix
\be
\J_{PC} = \J_{1/4}(\theta_3) \cdot \J_{1/2}(\theta_2)\cdot \J_{1/4}(\theta_1)
\ee
and selected the $\theta_i$'s to assure not having the identity or other
trivial matrix.

We varied the difference of $\theta_2$ between transmitter and
receiver over $\pi/15 \le \theta_2 \le \pi/2$.  We consider both
standard (equal within one part in $10^5$) and unequal ($\approx 5\%$
difference) absorption coefficients, $R_x$ and $R_y$. This is an
important distinction because in our calculations we found that
unequal absorption in the two polarization causes the light throughout
the ring to rapidly become uniformly polarized. By breaking this
symmetry between polarizations, the laser tends to approach a
preferred polarization direction. Therefore, it is important to
examine whether the synchronization behavior varies drastically with
regards to equal/unequal absorption.

\begin{figure}
\centerline{\psfig{file=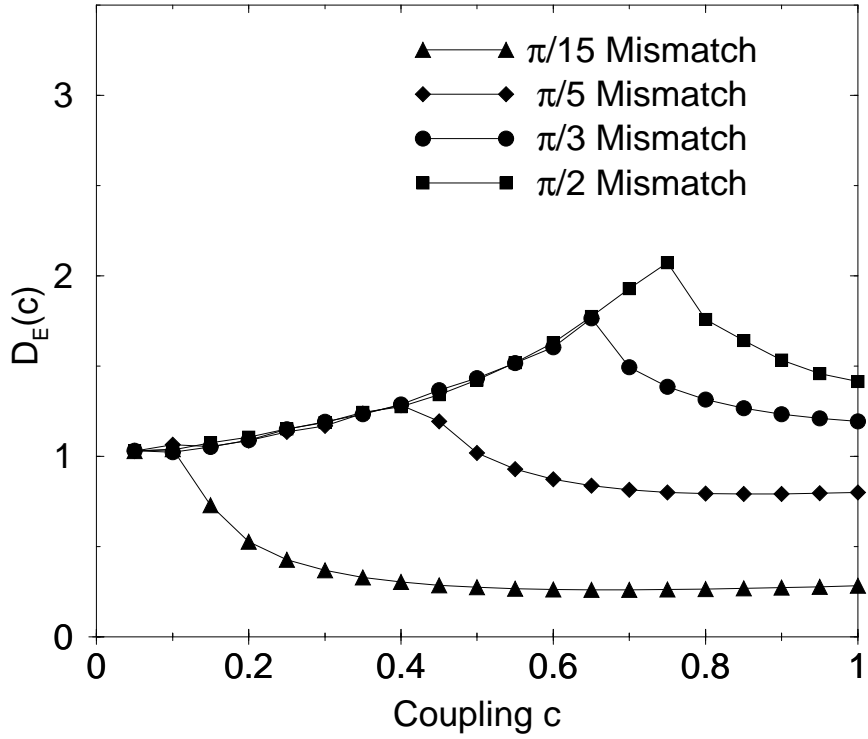,width=4.5in,angle=270}}
\caption{Identical synchronization error $D_E(c)$ for DFRLs with $\theta_2$ mismatch. 
There is nearly equal absorption, $R_x \approx R_y$. The DFRLs are first coupled for 
$20,000\tau_R$ and then the error term is averaged over an additional $3,000\tau_R$.}
\label{fig:amprms_jonesequal}
\end{figure}

We start by looking at the effect of deviations of $\theta_2$ on the
identical synchronization measure $D_E(c)$ case for almost equal
absorptions in Fig. \ref{fig:amprms_jonesequal}. The effect on
identical synchronization here is drastic. Even the small mismatch
($\pi/15$) has error above 10\% for all couplings. With the gain and
pump mismatches we saw small effects on identical synchronization,
here we see a very detrimental impact for even small mismatches.

\begin{figure}
\centerline{\psfig{file=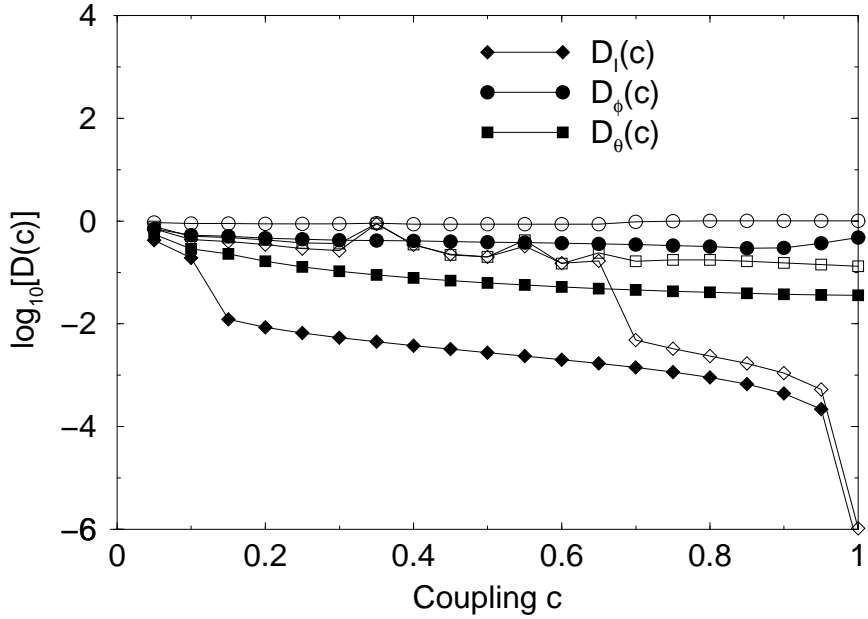,width=4.5in,angle=270}}
\caption{Generalized synchronization measures for DFRLs with $\theta_2$ mismatch. 
There is nearly equal absorption, $R_x \approx R_y$. The filled symbols represent
a mismatch of $\theta_2 = \pi/15$ and the clear symbols 
$\theta_2 = \pi/3$. The DFRLs are first coupled for $20,000\tau_R$ round trips and 
then the error terms are averaged over an additional $3,000\tau_R$.}
\label{fig:general_jonesequal}
\end{figure}

However, a look at the generalized intensity synchronization standard
deviation in Fig.
\ref{fig:general_jonesequal} shows some relatively strong synchronization. 
We see an odd sort of trend here. For each case, starting from $c=0$
the intensity synchronization error is unity, then at some value of
coupling strength, the intensity error drops quickly to $10^{-2}$,
representing about 1\% of the uncoupled normalization value. The
coupling value where this drop occurs systematically increases with
the $\theta_2$ mismatch value. Past this point, a generalized
synchronization error of $\approx 1\%$ may be exploitable for ASK, so
hope is not lost for success in the presence of polarization evolution
mismatch. Examination of the other generalized measures in Fig.
\ref{fig:general_jonesequal} show conclusively that with polarization
evolution mismatch, there is substantial loss of generalized phase and
polarization synchronization.

We then turn to the case of unequal absorptions, using absorption
coefficients $R_x = 0.45$ and $R_y = 0.425$. Although we might expect
different synchronization behavior with the different absorptions, we
find that the behavior is much the same as in the equal absorption
case in Fig. \ref{fig:amprms_jonesequal}. Similarly, looking at the
generalized synchronization measures, we again saw similar behavior to
the equal absorption case, with a notable similarity being good
synchronization values for the generalized intensity synchronization
measure.

Thus for the case of unequal absorptions, we still see a possibility
of utilizing ASK communication techniques. As stated above, this is
valuable information. For various optical modulation effects which we
may chose to use experimentally for communications, the light in the
ring must first be polarized. This is effectively what we have done by
setting the absorptions very unequal to each other. Thus, the fact
that the mismatch behavior was not very different leads us to
conjecture that a polarized light beam will possess similar
synchronization behavior to the general-case elliptically polarized
light we investigate through this study.

\subsubsection{Phase Mismatch}

Definitely the most foreboding experimental obstacle in trying to
synchronize two optical systems is matching the optical phase of the
two systems. The section above examining an identical transmitter and
receiver assumes that the optical phase between the two lasers can be
matched with perfect accuracy. Unfortunately, this is not a reasonable
assumption. First, the lengths of fiber in both rings would need to be
measured, cut and spliced together with an accuracy of a fraction of
the light's wavelength, $\approx 1.5
\mu$m. Second, complete phase stability would have to be achieved
between the two lasers. An EDFRL is always adapting its lasing mode to
achieve the mode of maximum power (minimum loss)~\cite{private}. This
causes the EDFRL to change its lasing wavelength across a wide
bandwidth on short time scales. Unless the lasers are somehow coerced
into identical lasing mode transitions, the phase stability between
the lasers will be poor which could be detrimental to synchronization
for $c<1$.

To model this kind of randomly changing phase shift, we consider a 
phase difference between the two lasers $\Delta\phi(t)$, which begins 
as some initial phase difference $\Delta\phi(t=0)$, and then on a 
time scale $\tau_{\phi}$, is shifted by a random phase amount. We take
\be
\Delta\phi(t+\tau_{\phi}) = \Delta\phi(t) +  \Delta\phi_{random}(\tau_{\phi}).
\ee
The coupling between the lasers is modified to be
\be
\E_R' = c\E_T + (1-c)\E_R e^{i\Delta\phi}.
\ee

\begin{figure}
\centerline{\psfig{file=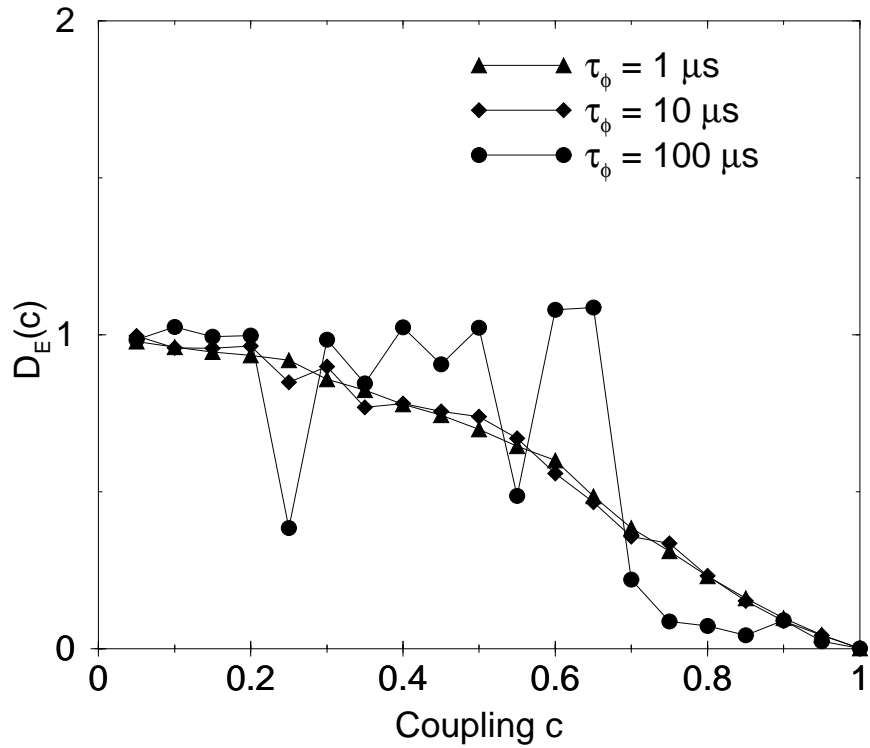,width=4.5in,angle=270}}
\caption{Identical synchronization error $D_E(c)$ for DFRLs with phase
mismatch. The lasers are otherwise identical. The DFRLs are first coupled 
for $20,000\tau_R$ and then the error term is averaged over an additional 
$3,000\tau_R$.}
\label{fig:phase_amprms}
\end{figure}

We look at the effect of phase shifts over a large range of time scales, 
from $1-100 \mu s$. The identical synchronization measure $D_E$ is plotted
in Fig. \ref{fig:phase_amprms}. With the exception of the long time scale 
($\tau_{\phi} = 100\mu$s), there is an orderly gradual progression from 
synchronization error of order unity for small $c$ to a small error for
large $c$. This simply tells us that the phase mismatch
has less effect for larger coupling where less of the phase mismatched
field from the receiver is being mixed with the field from the transmitter. 

\begin{figure}
\centerline{\psfig{file=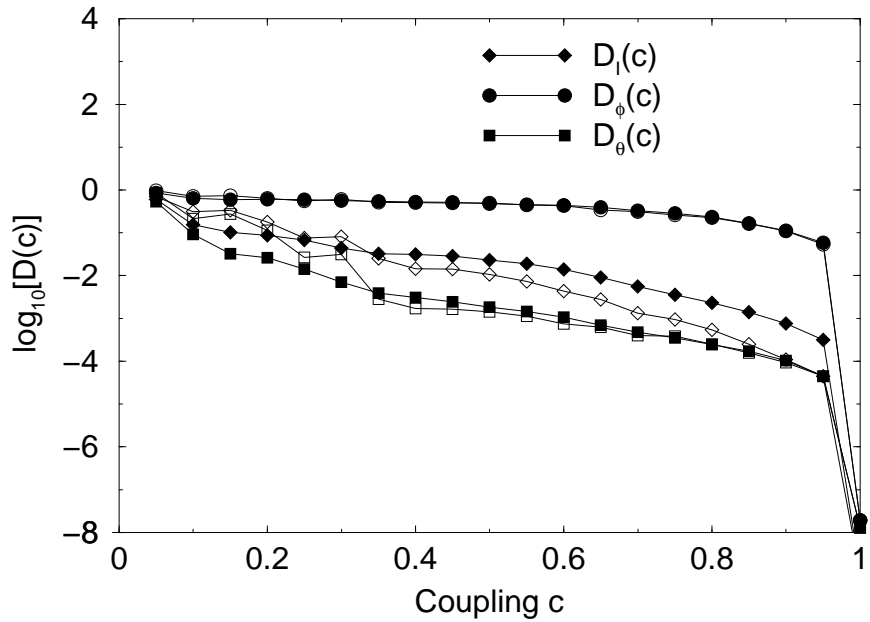,width=4.5in,angle=270}}
\caption{Generalized synchronization measures plotted against coupling constant c. 
There is phase mismatch. The lasers are otherwise identical. The filled symbols
represents $\tau_{\phi} = 1\mu s$ and the clear symbols $\tau_{\phi} = 10\mu s$.
The DFRLs are coupled for $20,000\tau_R$ and the measures are taken over an 
additional $3,000\tau_R$.}
\label{fig:phase_general}
\end{figure}

Looking at the generalized intensity synchronization measure $D_I(c)$ in 
Fig. \ref{fig:phase_general}, we see that the picture is not as bad. By 
a coupling of about $c=0.5$ we see that $D_I(c)$ is down to $10^{-2}$ and
keeps decreasing from there. Again, there may be enough intensity 
synchronization to communicate via ASK.

Moving now to phase synchronization $D_{\phi}(c)$ again in Fig. 
\ref{fig:phase_general}, we see qualitatively similar behavior to
$D_E(c)$. It appears that there is not a  good region of generalized phase 
synchronization until we are near $c=1$, which in 
effect is removes the problem of phase mismatch by removing the 
receiver's phase altogether. 

Polarization synchronization $D_{\theta}(c)$ in Fig. 
\ref{fig:phase_general} is relatively robust, the roving phase mismatch 
not contaminating the states of polarization. Like $D_I(c)$, we see that for
coupling strengths above $c \approx 0.5$, the synchronization error is
well under $10^{-2}$ and shows promise for communicating via polarization
modulation. 

\subsubsection{Length Mismatch}

Next we turn to synchronization with substantial mismatches in the
lengths of fiber in the two lasers. In the previous section, the
length mismatches were on the scale of the optical wavelength, so
the only complication was a mismatch in the phase of the two
fields. The slowly varying complex field envelopes were still
considered to be coincident. Now we look at mismatches much greater
than the general coherence length of the lasers. The amount of
mismatch which could be involved is dictated by the simulation. One
iteration time step in the simulation is equivalent to about $80$ ps
($\approx 1.7$ cm of passive fiber), so this is the smallest
macroscopic length mismatch we could examine.

The length mismatch on this order obliterated all
synchronization. However, this result isn't surprising. Once the
lasers are coupled, the resulting field is propagated around the ring
of the receiver and meets up with the incoming field from the
transmitter to be coupled and propagated once more.  If the length
mismatch is severe enough so that the receiver's complex envelope does
not match up with its corresponding evolved twin complex envelope
incoming from the transmitter, there is no reason to expect that there
would be any synchronization, since spatial envelope dynamics are only
very weakly correlated through the population inversion and the fiber
dispersion and birefringence effects. Thus, one evolving complex
envelope would be coupled to a different incoming complex envelope at
each trip around the ring.

Since the complex envelope propagates with a different round trip
time, we investigate the possibility that the lasers synchronize, but
with a time lag equal to the time it takes the light to travel the
length of the mismatch.  The possibility of robust synchronization,
but with a time lag caused by the substantial passive fiber length
mismatch prompts one last measure of generalized synchronization,
commonly called {\em lag synchronization}~\cite{avr,kocpar}. In the
context of our identical synchronization definition in equation
(\ref{eqn:identsynch}), we define a lag synchronization measure:
\be
D_E(c,\tau_{lag}) = \frac{1}{\Nc_E(\tau_{lag})}\left(\int_{K\tau_R}^{(K+M)\tau_R}H_E^2(c,t,\tau_{lag})dt\right)^{1/2},
\ee
where $\Nc_E(\tau_{lag}) = D_E(c=0,\tau_{lag})$ and
\be
H_E(c,\tau_{lag}) = \frac{|\Ec_T(t) - \Ec_R(t+\tau_{lag})|}{\avabs{\Ec_T}_{\tau_R}}.
\ee
This is basically a measure of the presence of identical
synchronization, just shifted in time by $\tau_{lag}$ between the
transmitter and the receiver. We can then proceed and similarly
include a time lag $\tau_{lag}$ in all of our generalized
synchronization measures, also.

\begin{figure}
\centerline{\psfig{file=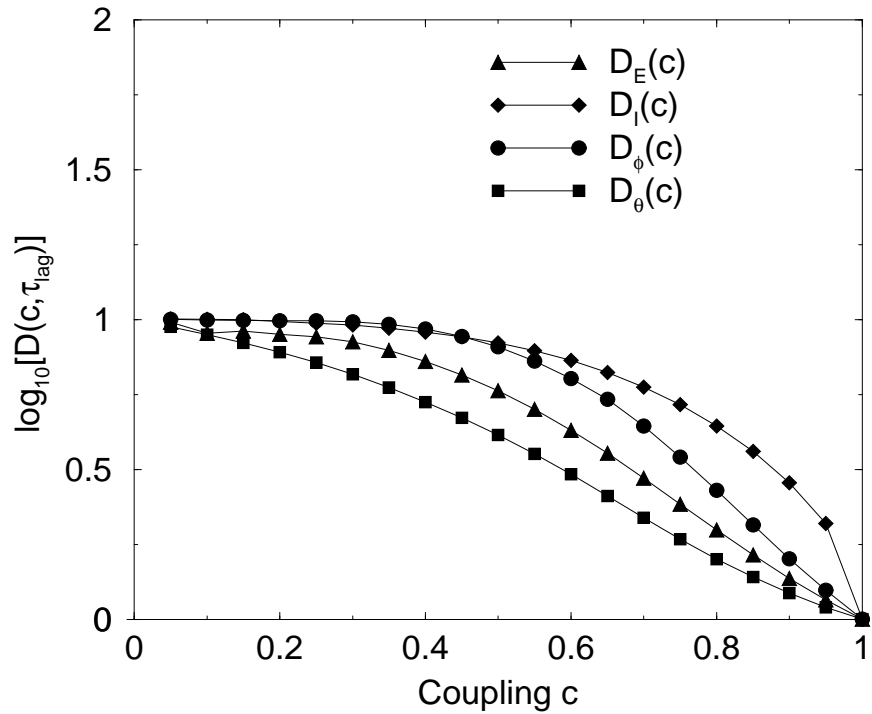,width=4.5in,angle=270}}
\caption{Synchronization errors for DFRLs with a length
mismatch of $\approx 80 \mathrm{ps}$. The lasers are otherwise identical. 
The DFRLs are first coupled for $20,000\tau_R$ 
and then the error terms are averaged over an additional 
$3,000\tau_R$.}
\label{fig:length_general}
\end{figure}

First we consider the case of identical synchronization. The plot of
$D_E(c)$ for this length mismatch case is plotted in Fig.
\ref{fig:length_general}.  
If we look for synchronization with a time lag of $\tau_{lag} = 80$
ps, we see behavior similar to the phase mismatch case where the
synchronization error starts at unity for small $c$ and gradually
descends toward synchronization for very strong coupling ($c \to 1$).
Examining the generalized synchronization measures for the $\tau_{lag}
= 80$ ps case in Fig. \ref{fig:length_general} we see that $D_I(c)$,
$D_{\phi}(c)$, and $D_{\theta}(c)$ rather much follow the pattern of
$D_E(c)$ with strong synchronization for $c\to 1$.

The overall result is that if large length mismatches are unavoidable or 
difficult to reconcile, suitable lag synchronization can be achieved 
through accounting for the time lag and using only very strong coupling 
strength values ($c\to1$).   

\subsection{Noise in the Communications Channel}

Two physical lasers will not have identical operating parameters, and 
synchronization of two physical lasers will unavoidably be subjected 
to the parameter mismatches just discussed. Another unavoidable issue is the effect on synchronization of noise in the channel by which 
the two lasers are coupled. Any physical application of synchronizing 
DFRLs will invariably be effected by this noise, so we examine that 
case here.

We consider signal to noise ratios 0 dB and 40 dB. We concentrate on the
case of two lasers with identical parameter values which are coupled via
a noisy fiber channel. The average noise amplitude $\avabs{\xi}$ we use is
determined from the signal-to-noise ratio given by
\be
SNR = 20 \log_{10}\frac{\avabs{\Ec_T}}{\avabs{\xi}}
\ee

Noise was added to the field arriving from the
transmitter before coupling. Instead of receiving as input $c\Ec_T +
(1-c)\Ec_R$, the receiver now receives the modified noisy input given by
$c\Nc_{\xi}(\Ec_T + \zeta) + (1-c)\Ec_R$, where $\zeta$ is a complex
polarization noise two-vector. The components of $\zeta$ are 
random Gaussian numbers with a standard deviation of 1, multiplied by 
the average noise amplitude $\avabs{\xi}$. The normalization on the incoming 
transmitter field plus noise is chosen so that the variance of this 
incoming `noisy' signal $\Nc_{\xi}(\Ec_T + \zeta)$ was equal to that of the clean transmitter field $\Ec_T$: 
\be
\Nc_{\xi} = \frac{\avabs{\Ec_T}}{\sqrt{\avabs{\Ec_T}^2 + \avabs{\xi}^2}}
\ee
Again the lasers were allowed to couple for 20,000 $\tau_R$ 
and then the error value $D_E(c)$ was averaged over the next 3,000 $\tau_R$.

\begin{figure}
\centerline{\psfig{file=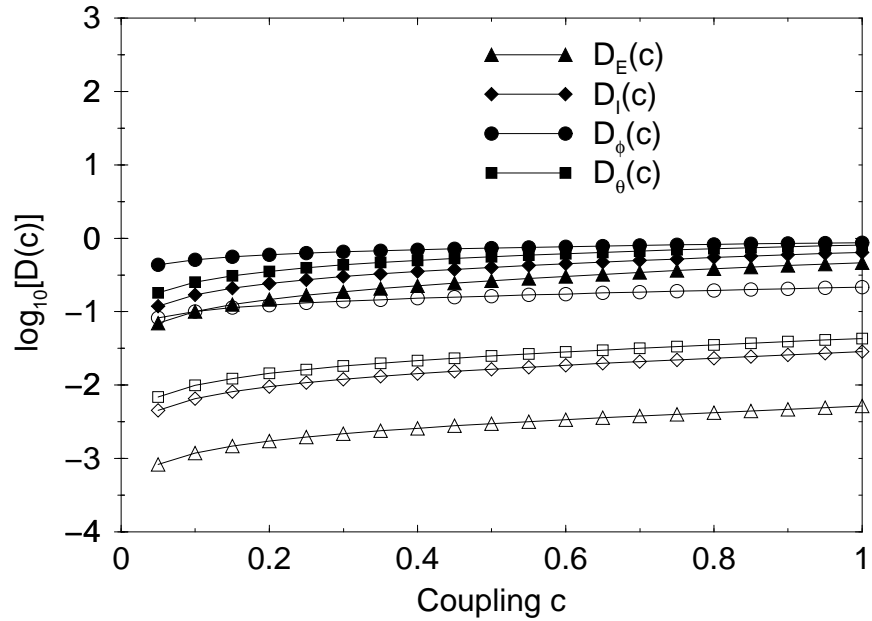,width=4.5in,angle=270}}
\caption{Synchronization errors for identical DFRLs with channel noise. 
The filled symbols represent SNR = 0 dB and the clear symbols SNR = 40 dB. 
The DFRLs are first coupled for $20,000\tau_R$ and 
then the error terms are averaged over an additional $3,000\tau_R$.}
\label{fig:noise_general}
\end{figure}

Looking first at $D_E(c)$ for identical lasers in Fig.
\ref{fig:noise_general}, we see that due to the noise there is a
steady growth in the synchronization error as coupling is
increased. Even for a SNR of 0 dB, at small coupling, $c
\le 0.1$, we have below a 10\% normalized RMS error. However, for large
coupling constants, the 0 dB SNR value leads to synchronization errors
of 20\% and more. For SNR of 40 dB, we see that for all
coupling constants the RMS error is well below a few parts in 1000. 
Here it is obvious that the rate of growth of the normalized RMS error 
as a function of coupling constant is very much the same for the 
range of SNR values. The SNR we quote is the channel signal to noise 
ratio, while the noise entering the receiver is $c\zeta$ when
we feedback $(1-c)$ of $\Ec_R$, thus the effective signal to noise ratio
in the receiver is higher when c increases from zero.

All the generalized synchronization measures in Fig.
\ref{fig:noise_general} show rather much the same behavior as the identical
synchronization case. For high SNRs, the synchronization is good for
all measures (except the phase synchronization measure), and all
measures continue to show a preference for weaker coupling, a type of
nonlinear noise reduction. In optical fiber systems, channel noise is
extremely low, and is usually not a concern.  Our purpose here is to
also examine what substantial noise, say due to multi-user
communications in the background, might do to our synchronization.  It
is encouraging that the lasers actually synchronize better for weaker
coupling in the presence of noise, as this might be a clue as to how
to utilize multiple channels for chaotic laser communication. This
fact matches well with our earlier observation that the most rapid
convergence of the two lasers into synchronization also occurs at
extremely low coupling. These two facts can perhaps combine in a
useful way later when examining communication methods more closely.

\subsection{Alternate Coupling Scheme}

The bulk of the synchronization problems which we encountered above
were due to the fact that the electric field is a complex,
polarization two-vector. Phase mismatch and fiber length mismatch were
found to be substantially detrimental to synchronization, as well as
mismatches in the evolution of the state of polarization for large
regions of coupling strengths.  This leads us to propose another way
to synchronize the lasers without coupling the full optical field of
the transmitter laser into the fiber ring of the receiver laser. We
examined a synchronization scheme where the electric field intensity
of the transmitter laser is detected, and used to electro-optically
modulate the optical intensity in the receiver laser in an effort to
drive the receiver into a state of generalized intensity
synchronization.

\begin{figure}
\centerline{\psfig{file=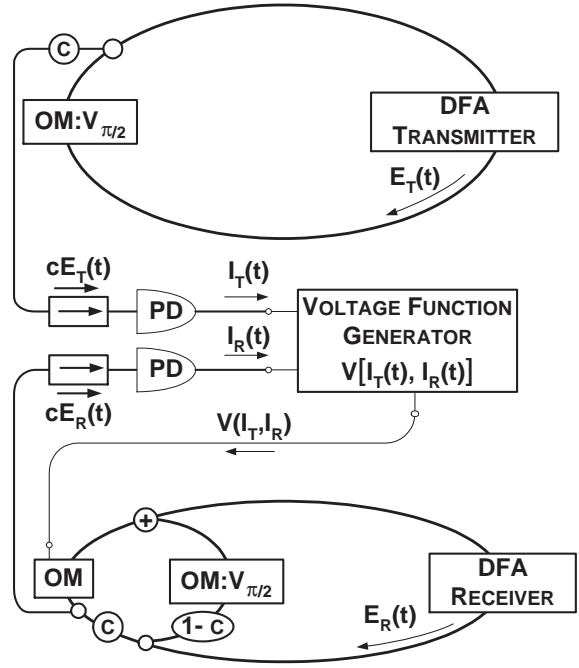,width=3in}}
\caption{Diagram of optical modulation coupling scheme. All optical modulators (OM)
are biased to a voltage of $V_{\pi/2}$. The optical fields of the transmitter and
receiver are detected by photodiodes (PD) and fed into a voltage function
generator. This voltage is then used to electro-optically modulate the $c$
branch of the receiver's ring to bring the receiver DFRL into a state of generalized intensity synchronization with the transmitter DRFL's intensity.}
\label{fig:modulator}
\end{figure}

In figure \ref{fig:modulator} the proposed intensity synchronization strategy is
diagrammed. This scheme is close to the previous optical amplitude
coupling strategy, except that we now insert a electro-optical
intensity modulator. The electro-optic modulator uses the incoming
electric field to destructively interfere with itself, thereby
lowering the total intensity of the incoming state. Technically, it
can do this in various ways~\cite{chuang}, but we chose to simulate a
Mach-Zehnder waveguide modulator. In this type of modulation, the
incoming intensity is split 50/50 into two adjacent channels of equal
length. One channel is then phase delayed using the properties of an
electro-optic crystal. The two channels are then merged again. If
there has been a phase delay created between the two channels, then
the two merged channels will destructively interfere and the optical
intensity will decrease. The physics behind such modulators is readily
found~\cite{yariv,chuang}, but the important result is that the amount
of phase shift between the two channels in the receiver $\Psi_R$ is
linearly proportional to the voltage applied across the electro-optic
crystal. We write this phase shift following the conventions
in~\cite{yariv} as
\be
\Psi_R = \pi\frac{V}{V_{\pi}}
\ee
where $V_{\pi}$ is the voltage needed to create a phase shift of
magnitude $\pi$. The net effect on the incoming intensity is
\be
\label{eqn:neteffect}
|\Ec^\prime_R (t)|^2 = \cos^2\left(\frac{\Psi_R}{2}\right)|\Ec_R(t)|^2,
\ee
where the primed (unprimed) field corresponds to the field after
(before) the electro-optic modulator.

The next step is to notice that due to the functional form of
equation (\ref{eqn:neteffect}), the receiver intensity can only be
modulated to a value of lower intensity. This would greatly hinder
synchronization as the control over $|\Ec_R|^2$ would be weakened.
For this reason, we put an identical electro-optic modulator in the
transmitter ring and bias that modulator with voltage $V_{\pi/2} =
V_{\pi}/2$ which gives a constant phase shift of $\Psi_T=\pi/2$. Then
according to the modulator equation for the transmitter which
corresponds to equation (\ref{eqn:neteffect}),
\bea
\label{eqn:trans_int}
|\Ec^\prime_T (t)|^2 & = & \cos^2\left(\frac{\Psi_T}{2}\right)|\Ec_T(t)|^2 \\
               & = & \frac{1}{2}|\Ec_T(t)|^2,
\eea
so the effective loss in the transmitter is $1/2$. This is equivalent
to increasing the absorption constants $R_x$ and $R_y$ to include a
$50\%$ intensity loss to the transmitter. If we now bias the
electro-optic modulator in the receiver to a unmodulated state of
$\Psi_R=\pi/2$, then the lasers are again identical, and
synchronization conditions are favorable. The reason for doing this is
now the receiver's intensity can be modulated upwards by lessening the
phase shift from the unmodulated state $\Psi_R=\pi/2$.

Keeping these biasing concerns in mind, we now describe the
synchronization scheme.  We detect the incoming transmitter electric
field with a photodiode to create a current proportional to
$|\Ec_T|^2$.  Meanwhile, the receiver intensity is detected before the
modulator by another photodiode and a current proportional to
$|\Ec_R|^2$ is also created. These currents are input into a voltage
function generator which outputs a voltage to the electro-optic
modulator. We note here the considerable physical task required as all
these propagation times must be matched appropriately. Here we assume
that we can physically create the ideal voltage function:
\be
V(|\Ec_T(t)|^2,|\Ec_R(t)|^2) = \frac{2V_{\pi}}{\pi}\cos^{-1}\left(\frac{|\Ec_T(t)|}{\sqrt{2}|\Ec_R(t)|}\right).
\ee
This will give a phase shift of
\be
\Psi = 2\cos^{-1}\left(\frac{|\Ec_T|}{\sqrt{2}|\Ec_R|}\right).
\ee
Using this phase shift in equation (\ref{eqn:neteffect}), we see that we 
immediately arrive at  intensity synchronization, because
\bea
|\Ec^\prime_R (t)|^2 & = & \cos^2\left(\cos^{-1}\left[\frac{|\Ec_T(t)|}{\sqrt{2}|\Ec_R(t)|}\right]\right)|\Ec_R(t)|^2 \\
               & = & \frac{1}{2}|\Ec_T(t)|^2,
\eea
which is exactly equal $|\Ec^\prime_T (t)|^2$ in equation (\ref{eqn:trans_int}) in its biased state. One hindrance
we must keep in mind is that we can not allow voltage functions
where $|\Ec_T(t)|^2 > 2|\Ec_R(t)|^2$ because then the arc cosine argument
will be greater than one. Therefore we limit
the voltage by imposing the condition that if $|\Ec_T(t)|^2 > 2|\Ec_R(t)|^2$,
then 
\be
V(|\Ec_T(t)|^2,|\Ec_R(t)|^2) = \frac{2V_{\pi}}{\pi}.
\ee
This may cause a slight delay in original synchronization, but once the
intensities are close to synchronization, it will not be a factor at all.

We also allow for a variation of coupling constants by splitting the
receiver ring into two branches in the proportion $c:(1-c)$. The
branch $c\Ec_R$ then goes through the electro-optic modulator, and the
$(1-c)\Ec_R$ diverts around the modulator and is subjected to a 50\%
intensity attenuation. The two branches are then joined again before
entering the DFA.

\begin{figure}
\centerline{\psfig{file=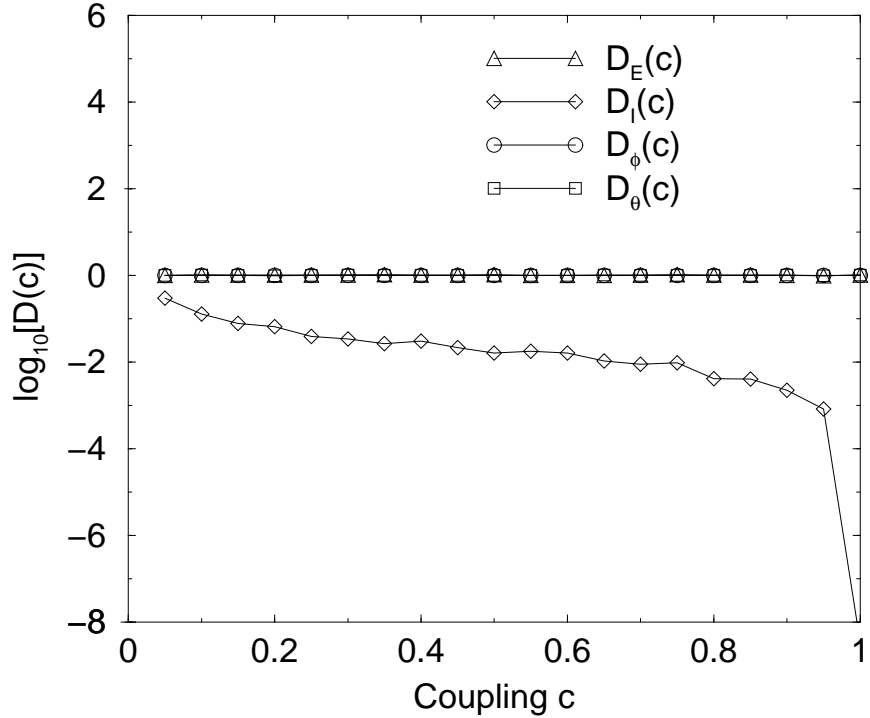,width=4.5in,angle=270}}
\caption{Synchronization errors for identical DFRLs with 
coupling by an optical modulator. The DFRLs are 
first coupled for $20,000\tau_R$
and then the error terms are averaged over an additional $3,000\tau_R$.} 
\label{fig:mod_general}
\end{figure}

The results are shown in Fig. \ref{fig:mod_general}.  Since only the
intensity is synchronized, there is no sign of identical
synchronization. However, we see that for larger $c$, there is good
{\em generalized intensity synchronization} within $10^{-2}$.  We will
attempt to communicate via ASK with this configuration and discover
whether or not intensity synchronization with errors of order
$10^{-2}$ is good enough for reliable message recovery. Figure
\ref{fig:mod_general} shows there is no phase synchronization,
there is no reason to expect otherwise. 
The same is true for the polarization synchronization. This also
is to be expected, since  
no polarization information is shared between the 
lasers.

All in all, it seems like relatively successful generalized intensity
synchronization can be achieved through optical modulator
coupling. However, there are some unanswered questions. One is that
the photodiodes which detect the intensities have finite bandwidths
(up to the order of GHz). The question here is whether the lasers
synchronize with just lower frequency information being shared from
transmitter to receiver? Also, what the possibilities for a function
of the form $V(|\Ec_T|^2,|\Ec_R|^2)$? Are there physically high-speed
functions which maximize the efficiency of the synchronization?  These
and other questions will be the focus of more study~\cite{synchmod}.

\section{Communications}

\begin{figure}
\centerline{\psfig{file=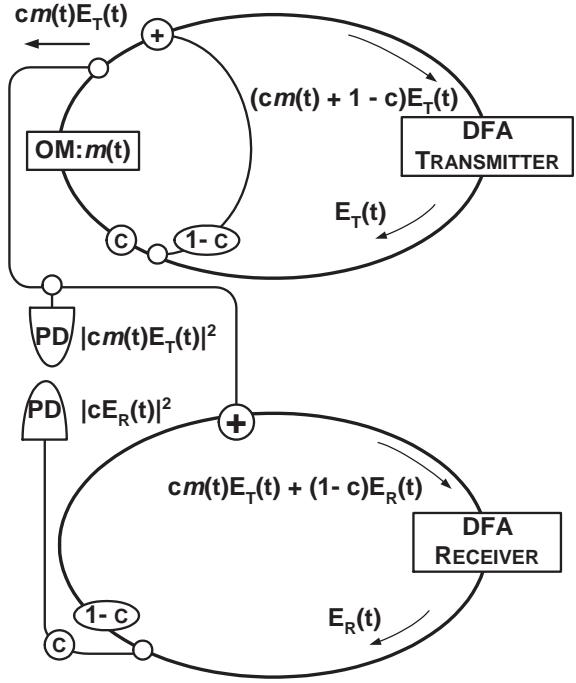,width=3in}}
\caption{Diagram of communications using optical coupling scheme. The setup
is almost identical to Fig. \protect\ref{fig:scheme} except it contains and additional 
$c:1-c$ branch in the transmitter. This branch must be included if synchronization
is to be achieved for couplings in the range $0 < c < 1$. Note for $\Ec_T(t)=\Ec_R(t)$,
the optical fields entering both active mediums are identical, even with the presence
of modulation. }
\label{fig:comm_optical}
\end{figure}

We will attempt transmission and recovery of a bit string using a
simple electro-optically modulated ASK technique for identical lasers
with optical coupling and with coupling by electro-optic modulation.
This is the main method employed in the experimental cases by Roy and
VanWiggeren ~\cite{science,prlrvw,private}.  The set up is shown
schematically in Fig. \ref{fig:comm_optical}.

An electro-optical modulator must be added to the transmitter ring in
order to electro-optically modulate the bit string onto the chaotic
optical waveform in the transmitter. One method would be to just take
the scheme in Fig. \ref{fig:scheme} and modulate the transmitter
optical intensity. However, we must look closer at the effect of this
scheme upon the state of synchronization.

As noted throughout the section on synchronization, the mean round
trip intensity of an EDRFL is a relatively constant value in the long
time asymptotic regime. As a message is modulated onto the optical
intensity of the transmitter, we must take care to keep the mean round
trip optical intensity constant, or we will send the laser back into a
regime of chaotic transients and the robust, steady-state behavior of
the EDFRLs will invariably be lost. Therefore, we must modulate in a
manner which retains the magnitude of the mean intensity. Our choice
is to modulate the intensity up a certain value $K$ for a ``1" bit,
and modulate down the intensity by $K$ for a ``0" bit.  If the bit
values are equally probable, then the long term mean intensity will be
retained. We define a message parameter $m(t)$ to be a variable
multiplicative value on our intensity such that the encoded intensity
value is $|m(t)\Ec_T(t)|^2$ where,
\be
m = \sqrt{1+K} \mbox{        for a ``1" bit},
\ee
and
\be
m = \sqrt{1-K} \mbox{        for a ``0" bit}.
\ee
For the following simulations we used a value of $K=0.1$.

If we now insert the electro-optic modulator into the transmitter fiber
ring and modulate the optical intensity by $m$, we will run into 
synchronization problems for $c \ne 1$ as discussed now. 
The integrated population inversion
in the active medium is nonlinearly dependent upon the incident
optical intensity (see equation (\ref{eqn:popinv})). We see that the 
incident optical intensity in the transmitter will equal
\be
|\Ec_T(t+1)|^2 = m^2(t)|\Ec_T(t)|^2.
\ee
However, when the transmitter optical field is coupled into the receiver
ring, it is effectively multiplied by the coupling strength $c$ and added to
the receiver's optical field multiplied by $(1-c)$. The resultant
intensity seen by the receiver's active medium is
\be
|\Ec_R(t+1)|^2 = |cm(t)\Ec_T(t) + (1-c)\Ec_R(t)|^2.
\ee
Even if the lasers begin in perfect synchronization, i.e.,
$\Ec_R(t)=\Ec_T(t)=\Ec(t)$, synchronization will be lost by 
modulating a message since the active media will see
different intensities
\bea
I_T(t) & = & m^2(t)|\Ec(t)|^2 \\
I_R(t) & = & (cm(t) + 1 - c )^2|\Ec(t)|^2.
\eea
For $c=1$, this is not a problem (as was experimentally
shown in~\cite{science,prlrvw,private}), but for $c \ne 1$
synchronization will be lost.

To address this problem we introduce the idea of partial modulation of the
transmitter intensity. In the transmitter fiber ring, before the
intensity is modulated, we split the optical field in a proportion
$c:(1-c)$ (Fig. \ref{fig:comm_optical}). 
The branch corresponding to the $c$ value is electro-optically
modulated, and the $(1-c)$ branch is not altered. Before the two branches
are rejoined, the modulated field is coupled out of the $c$ branch and
sent off to the receiver. Therefore, if the lasers are in synchronization
their active media both see the same optical intensity
\be
I(t) = |cm(t)\Ec(t) + (1-c)\Ec(t)|^2.
\ee
Synchronization may persist in the presence of
message modulation, because the synchronized state is still a solution
to both the receiver's and the transmitter's dynamical system even in the
presence of arbitrary modulation.  Of course, the more difficult
question of stability remains.

\subsection{Identical Lasers with Optical Coupling}

We take this dual EDRFL system and transmit a message. We choose an
non-return-to-zero (NRTZ) bit rate of 1 GHz, corresponding to 
13 model integration iteration time steps per bit. We need to recover the 
bits at the receiver. The incoming intensity from the transmitter is
detected by a photodiode and produces a current proportional to the
value $|c m(t) \Ec_T(t)|^2$ (see Fig.
\ref{fig:comm_optical}). We simultaneously couple out the optical
field from the receiver ring with a $c:(1-c)$ coupler and detect the
value $|c\Ec_R(t)|^2$ with another photodiode. If
$\Ec_T(t)=\Ec_R(t)=\Ec(t)$, the transmitter's intensity divided by the
receiver's intensity will recover the message,
\be
\frac{|c m(t) \Ec(t)|^2}{|c \Ec(t)|^2} = m^2(t).
\ee
The overall decision on a ``0'' bit or a ``1'' bit is made over all $N$ time steps within the bit
time period by calculating the average $m^2(t)$ value received over
the $N$ time steps and making a decision using the rules
\bea
D_{bit} & = & 1 \mbox{   if  }\frac{1}{N}\mathop{\sum}_{i=1}^N m^2(t_i) \ge 1.0 \\
D_{bit} & = & 0 \mbox{   if  }\frac{1}{N}\mathop{\sum}_{i=1}^N m^2(t_i) < 1.0.
\eea
In our simulation $N = 13$ here, with $t_i$ corresponding to the $i$th integration time step of the model.

We transmit $10^7$ independent random bits and record the bit error rate
\be
\mbox{BER} = \frac{\mbox{errors recorded}}{\mbox{bits transmitted}}.
\ee
If no errors occur, we report a BER of zero, noting that this is
only true up to the first $10^7$ bits sent.  For couplings in the
range $0.005 \le c \le 1.0$, we obtained error-free recovery of the
whole bit string.  This is consistent with the above conjecture that
the modified coupling scheme will not lose synchronization with the
addition of intensity modulation. If synchronization is not being
effected, one could further conjecture that errors will never arise in
the long term state since synchronization will continue to be just as
robust. No claim of proof of this fact is made here, as it is conceivably
possible that for sufficiently deep modulation stability properties could
change. 

\begin{figure}
\centerline{\psfig{file=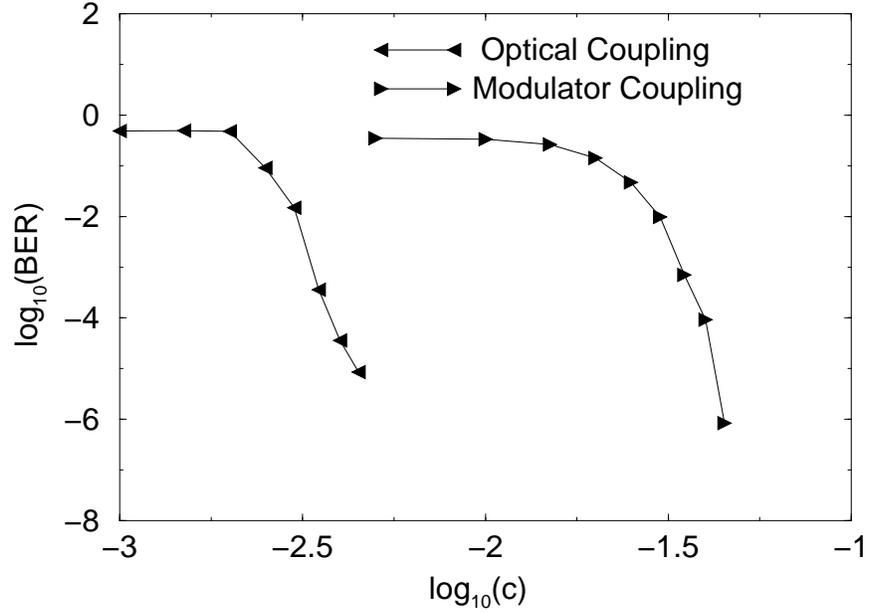,width=4.5in,angle=270}}
\caption{Bit error rate versus coupling for the case of identical lasers
using the two different coupling schemes. The coupling range is $ 0.0 \le c \le 0.06$. The encoding is done via ASK modulation at a bit rate of 1 GHz with modulation factor
$K = 0.1$.}
\label{fig:ber_small}
\end{figure}

We turn to a search of the minimum error-free coupling strength. In
Fig. \ref{fig:ber_small} we see that we begin to get nonzero BERs
below a critical coupling of $c_{crit} = 5.0\times 10^{-3}$. The
error-free recovery of bits for such small couplings is
remarkable. The coupling scheme practically guarantees this since the
lasers synchronize at such small coupling strengths to begin with. We
note a small difference in the critical coupling found for straight
synchronization in the previous section ($c_{crit} = 1.3\times
10^{-3}$), and the critical coupling for communications ($c_{crit} =
5.0\times 10^{-3}$).  It is possible that the electro-optic modulation
is actually increases the largest lyapunov exponent in the rings
(found above to be approximately $(6.3\pm 0.3)\times 10^{3} \mbox{
s}^{-1}$ without electro-optic modulation) which would then increase
the largest conditional lyapunov exponent, and thereby raise the value
of the critical coupling needed from the simple synchronization level
to the calculated level of necessary coupling strength needed for
synchronization in the communications case.

\begin{figure}
\centerline{\psfig{file=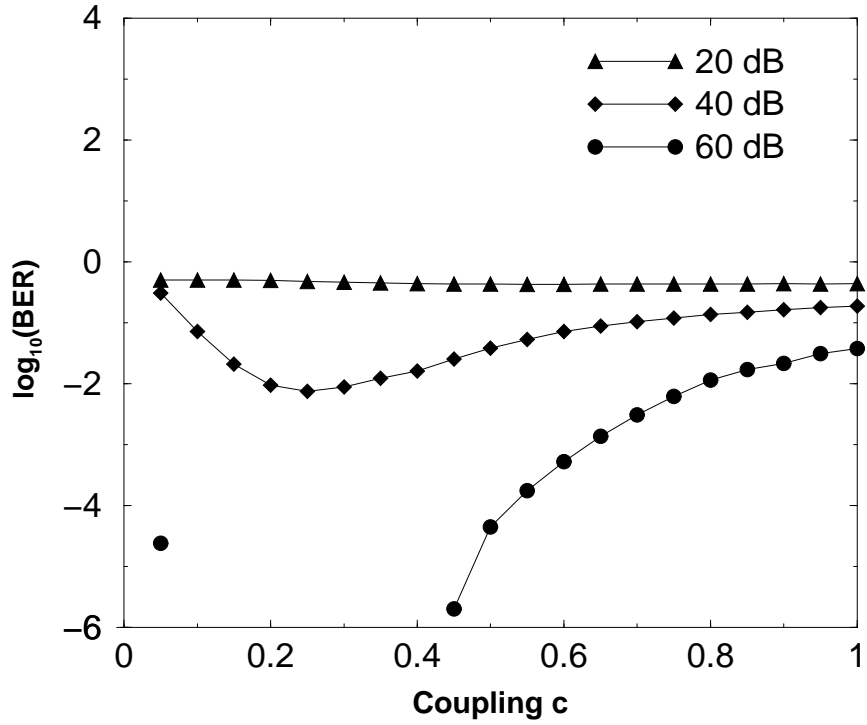,width=4.5in,angle=270}}
\caption{Bit error rate versus coupling for the case of optically coupled 
identical lasers in the presence of communication channel noise. The 
encoding is done via ASK modulation at a bit rate of 1 GHz with a modulation
factor of $K=0.1$.}
\label{fig:ber_noise}
\end{figure}

To complete the bit error rate calculations, we include the performance
of the optically coupled system when faced with communication channel 
noise. In Fig.~\ref{fig:ber_noise} we plot the BER versus coupling for 
SNRs in the range 20 dB to 60 dB. We see that for a 
SNR of 20 dB, there is no message recovery. In this case, the variance of
the channel noise is equal to the modulation amount ($K=0.1$) so lack of
recovery is not surprising. We see an improvement at a SNR of 40 dB, where a
range of lower coupling values are preferred. This improvement is accentuated at a 
SNR of 60 dB where the BER = 0 (up to $10^7$ bits) for couplings in the
range $0.1 \le c \le 0.4$, while with increasing coupling we get BERs on the
order of $10^{-2}$ as $c \to 1$.
These results further confirm the previous indications that for 
optically communicating with the method here described, better success 
may be achieved by using coupling strengths much less than $c=1$.

\subsection{Identical Lasers with Coupling by Electro-optic Modulation}

\begin{figure}
\centerline{\psfig{file=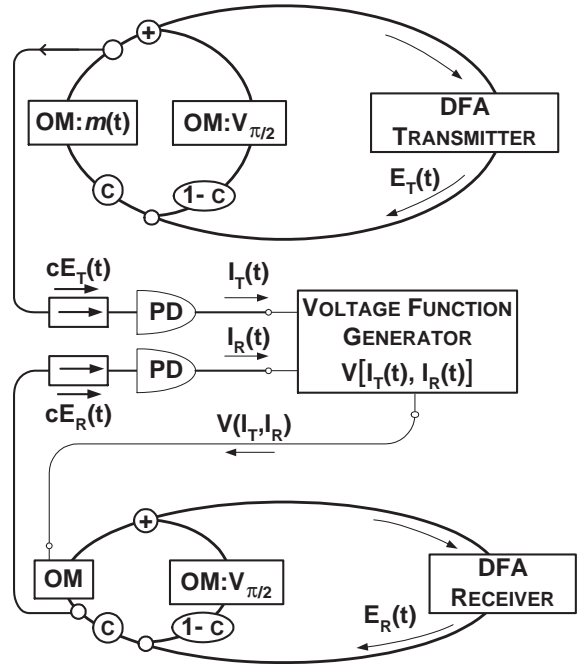,width=3in}}
\caption{Diagram of communications using optical modulation coupling scheme. Again,
the scheme is almost identical to Fig. \protect\ref{fig:modulator} except for the presence
of a $c:1-c$ branching in the transmitter. Again, this is added in an attempt to
achieve synchronization in the presence of modulation for couplings in the range
$0 < c < 1$.}
\label{fig:comm_modulator}
\end{figure}

We next modify our transmission scheme in the spirit of our
alternative method of coupling by electro-optic modulation. The set up
is similar to the above optical coupling set up except we now include
an identical $c:(1-c)$ fiber branching in the receiver laser which is
identical to the one in the transmitter laser (Fig.
\ref{fig:comm_modulator}). As before, this is included so that the
active medium population inversions in the two lasers, if already
intensity synchronized, will see the same incident intensities and
remain synchronized. We note here that we previously found that the
only robust synchronization in this method of coupling was generalized
intensity synchronization.  So this examination serves as a test of
whether or not ASK is feasible with only intensity synchronization
present.

Unlike the optical coupling method above we do not need to make
special provisions to recover the incoming encoded message. The
recovery method is already in place. Once the lasers are synchronized,
the voltage function generator will be putting out an relatively
constant voltage of $V_{\pi}/2$.  Once the message starts arriving,
this voltage function will respond in a manner to retain
synchronization. If a ``1" bit is transmitted, then the incoming
intensity value $|c m(t) \Ec_T(t)|^2$ will cause the voltage function
generator to decrease its voltage (thereby decreasing the phase shift
and raising the receiver's intensity). A likewise voltage increase
will occur if a ``0" bit is transmitted. Therefore we can just monitor
the voltage and make our overall decision variable over the $N$ time
steps within the bit time period as
\bea
D_{bit} & = & 1 \mbox{   if  }\frac{1}{N}\mathop{\sum}^{N}_{i=1} V_i \le \frac{V_{\pi}}{2} \\
D_{bit} & = & 0 \mbox{   if  }\frac{1}{N}\mathop{\sum}^{N}_{i=1} V_i > \frac{V_{\pi}}{2}.
\eea

Again we send $10^7$ random bits 
and calculate appropriate BER plots.
We found that the BER was more sensitive to the modulation
amount value $K$ that in the optical coupling case. This is
likely due to the less robust synchronization of this
method compared to the optical coupling method. 
By roughly optimizing the modulation amount value in the range 
$0.01 \le K \le 0.1$ for coupling strengths in the range 
$0.05 \le c \le 1.0$, we were able 
to achieve error-free recovery (again noting that
this is only accurate up to the first $10^7$ bits). The
existence of generalized intensity synchronization is sufficient
for suitable ASK message recovery.

Again we search for the critical minimum coupling strength for
error-free transmission in Fig. \ref{fig:ber_small} for 
coupling strengths in the range $0.0 \le c \le 0.05$. Here
we do see a critical coupling an order of magnitude higher
than the optical coupling case, even with the optimizing of the 
modulation amount $K$. The critical coupling strength
appears to be $c_{crit} \approx 4.5\times 10^{-2}$. This higher critical
coupling strength is again likely due to the fact that
the generalized intensity synchronization error is
more robust in identical lasers with optical coupling than
in identical lasers with coupling by electro-optic modulation.
However, the scheme has the potential for much improvement
and a more detailed study will follow in~\cite{synchmod}.

\section{Cryptographic Setting of our Work}

\begin{figure}
\centerline{\psfig{file=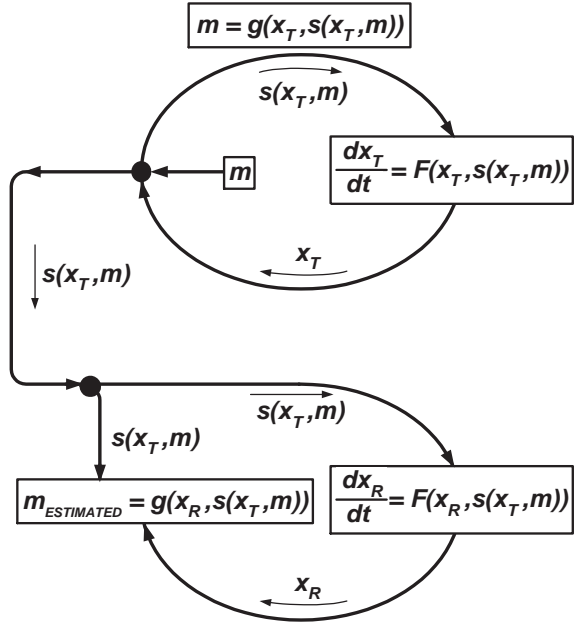,width=3in}}
\caption{Cryptographic Setting of our $c=1$ communications scheme. This is known as a ``Self-synchronous stream cipher in cipher feedback mode (CFB)''
\label{fig:cfb}}
\end{figure}

In the cryptographic literature~\cite{denning,kuhn} the communications
strategy investigated by us in this and earlier
papers~\cite{prlak,sush} and by Roy and
VanWiggeren~\cite{science,prlrvw} with $c=1$ is identified as a
self-synchronous stream cipher working in cipher feedback mode. The
block diagram we show in Fig. \ref{fig:cfb} captures our ideas in
a general format, and corresponds closely to Figure 3.9 of
Denning~\cite{denning} and Figure 2 of K\"uhn~\cite{kuhn}.

In our diagram the transmitter system is characterized by a state variable $\x_T$ satisfying the differential equation
\be
\frac{d \x_T}{dt} = \F(\x_T, s(\x_T,m)).
\ee
The state variable $\x_T$ is mixed with the message by an operation we
call $s(\x_T,m)$ which we assume has an inverse, so that $m$ can be
recovered by $m = g(\x_T,s(\x_T,m))$. In this paper we have
considered mixing realized by multiplication,
which has a simple inverse. Other invertible operations are also
possible.

The signal $s(\x_T,m)$ is transmitted out of the transmitter toward
the receiver as well as being fed back to a port in the nonlinear element of the
transmitter. At the receiver the signal $s(\x_T,m)$ enters the
nonlinear receiver element which is governed by the differential
equation for the state $\x_R$
\be
\frac{d \x_R}{dt} = \F(\x_R, s(\x_T,m)).
\ee
The output $\x_R$ of this nonlinear operation is then delivered to the
operation $g$ along with the received signal $s(\x_T,m)$, and the message is estimated as
\be
m_{estimated}=g(\x_R,s(\x_T,m)),
\ee
which is precisely $m$ when the systems are synchronized $\x_R = \x_T$.

K\"uhn notes that it is the function $\F(\x,s)$ in which the
cryptographic strength, if any, lies, and he outlines a method for
selecting the function so that it meets certain desired attributes
from a security point of view. He then constructs a function following
his guidelines. He notes that there are several levels of
cryptographic attack on this (or any) secure communications idea. In
each case one assumes that the transmitter and receiver are known in
detail to a cryptanalyst. The attacker must determine which
parameter settings in the transmitter and receiver nonlinear functions
are being used at the time the attack is made.
\begin{itemize}
\item {\bf chosen cleartext (or plaintext)} and ciphertext which the analyst can determine by putting the chosen plaintext through the transmitter. If, as K\"uhn assumes, there is a finite set of permissible nonlinear functions, this attack is straightforward in principle, though it make take an unacceptably long time to accomplish. K\"uhn gives an interesting estimate of the number of operations to attack his algorithm.

\item {\bf known cleartext (or plaintext)}. This differs from the previous situation in that the cryptanalyst may not choose arbitrary plaintext to pass through the transmitter, but is somehow restricted in the permitted messages.

\item {\bf ciphertext only}. This is the situation where the attacker observes a transmission, but does not know what the parameter settings in the nonlinear function were during the transmission. The cryptanalyst always has the option in this case of passing the observed ciphertext through the receiver and adjusting parameters in the receiver until discernible plaintext is created in $\hat{m}$.
\end{itemize}
K\"uhn notes that {\bf ciphertext only} is ``the usual
situation in practice."

Subsequent to K\"uhn's work several papers appeared indicating
successful attacks on the particular algorithm and on the class of
algorithm he introduced~\cite{heys,anderson,millan}. Before K\"uhn's
paper there appeared an interesting successful attack on stream
ciphers using ciphertext alone~\cite{sieg}.

Generally the cryptographic strength of the algorithm (read, nonlinear
function  used in  the  transmitter  and receiver)  is determined   by
specific cryptanalysis on each example  or on a  class of examples. An
excellent review of stream   ciphers is given  by Rueppel~\cite{ruepp}
who also notes that ciphers of  any sort are often compromised by key
mismanagement rather than the intrinsic strength of the cipher scheme.

As noted in the introduction, we do not provide a cryptanalysis of any
of our algorithms which in this paper are comprised of the physical
equations of motion of a rare-earth-doped ring laser. Perhaps the
cryptographic setting of this communications system will permit others
with capability in this area to provide such a cryptanalysis.  It is
interesting to note that our systems with $c \ne 1$ are not generally
found in the cryptographic literature, and perhaps their analysis will
provide new insights into communications security.

\section{Conclusions and Discussion}

We began with a study of the quality of synchronization when all the
parameters of the transmitter and receiver lasers were identically
matched.  Synchronization occurred rapidly for practically all
coupling strengths. The critical coupling strength, below which no
synchronization occurs, was found to be $c_{crit} = 1.3\times
10^{-3}$, for our parameters. For strong coupling ($c\to 1$), we found
that synchronization sets in essentially instantaneously ($\tau_S \le
1\mu$s). Upon examination of the temporal behavior of the
synchronization error, however, we found evidence of two distinct time
scales. There is an immediate jump towards synchronization due to the
initial mixing of the optical fields. The size of this jump is
proportional to $c$, the largest jump coming at $c=1$. After this
initial jump, a second rate of convergence takes over due to the
asymptotic relaxation of the population inversion in the active medium
to its equilibrium value. We numerically showed that this second
convergence rate has a maximum in the coupling range $0.02 \le c \le
0.04$, and decreases to a minimum rate at $c=1$. We analytically
demonstrated the global stability of the synchronization manifold at
$c=1$, and also determined a lower bound on the magnitude of the
convergence rate.

The extremely small value of c for which synchronization occurs in this erbium-doped laser model stems from the very long fluorescence time $T_1 \approx 10$ ms for erbium. This leads to the waveform in the fiber ring changing very slowly in time; in fact, changes take of order $10^5$ round trips. This means that even as we replace the waveform in the uncoupled receiver just a small amount at each round trip for small $c$, there is ample time to fill the receiver with the same waveform as in the transmitter. If we increase $T_1$ by hand (not physically) synchronization for such remarkably small coupling is no longer found.

We then turned to the examination of the effect of parameter mismatch
between the two laser subsystems. We defined several measures of
generalized synchronization in detail for the purpose of examining
whether certain measurable properties of the lasers remained in
synchronization as parameter mismatch was increased. The properties
examined for synchronization were the optical phase, field intensity,
state of polarization, and in one relevant case we calculated the time
lag synchronization.

We first looked at mismatches in the gains of the rare-earth-doped fiber
amplifiers and the pump levels of those amplifiers. Identical
synchronization quality was degraded somewhat, but for reasonable
mismatches we maintained good generalized synchronization in the
optical intensity and state of polarization measures. When we looked
at mismatches in the evolution of the states of polarization by
mismatching the fiber propagation Jones matrices, the effect was more
drastic. Identical synchronization was detrimentally impacted along
with the generalized optical phase and state of polarization
synchronization measures. However, depending on the size of the
mismatch, there were regions of relatively good generalized intensity
synchronization. Whether these regions are good enough to exploit for
communications is not obvious. We found that the synchronization
behavior was relatively unchanged as asymmetry was introduced in the
absorptions in the two polarizations. This would tend to imply that
polarized optical fields possess synchronization behavior similar to
their general-case elliptically polarized counterparts.

We then turned to mismatches in the lengths of the ring lasers. Both
very short length mismatches which cause a mismatch in the optical
phase between the two laser subsystems and large length mismatches
which cause completely different round trip times for the complex
envelope amplitudes in the two rings. For the case of optical phase
mismatch, all synchronization was severely effected for couplings for
$c \le 0.5$. For $c \ge 0.5$, we began to get lower synchronization
errors with increasing coupling strength for the intensity and state
of polarization measures. The indication would be that considering
phase mismatch, success could be achieved by staying in the strong
coupling regime only. 

This type of phase mismatch is commonly cited as the major barrier 
in achieving ``true'' optical synchronization, i.e., completely 
synchronized, coupled, entirely optical systems with $c < 1$. Therefore,
any serious chance for optical synchronization needs to address
the physical issue of optical phase mismatch. We suggest the following as
a possible line of attack on the problem. 

In their experimental
work at Georgia Tech, VanWiggeren and Roy~\cite{prlrvw} included
an examination of the passive ring structure consisting of two
fiber loops of different lengths. When the two loops are rejoined
the ring laser dynamics act to optimize the resulting intensity,
\be
|(\bfE_{R1} + \bfE_{R2})|^2 = (|\bfE_{R1}|^2+|\bfE_{R2}|^2 +
2|\bfE_{R1}||\bfE_{R2}|\cos\theta_R \cos\phi_R),
\ee
using the notation in~\cite{prlrvw}. Here, $\theta_R$ and $\phi_R$ are 
respectively, the angle between the states of polarization and the
phase difference between the two optical fields at the point where
they are rejoined. By optimizing the intensity, 
the cross term $\cos\theta_R \cos\phi_R$ goes to an almost fixed value.
Therefore, there is a certain amount of phase stabilization
occurring due to the optimization effect.

We theoretically studied one of these dual-ring systems in
\cite{akbl1} and subsequently have numerically observed this same 
cross term maximization. However, in~\cite{akbl1} we also reported on a type
of frequency filtering which occurs due to the two
different times of propagation through the two loops. This type of 
filtering causes the frequency spectrum to be less broadband and more resembling 
quasi-periodicity. 

However, if one took the two-loop configuration,
but made the lengths of the loops as identical as possible, there
would exist this cross term maximization, but without the
frequency filtering, assuming the difference in propagation
times through the two loops was less than the correlation
time of the slow-varying complex optical field envelope.
Hence, the advantageous stabilization of the optical phase
in the transmitter would still be present, but without
paying the penalty of frequency filtering. Perhaps
experimental work in the future will examine this idea further.

For large mismatches, all synchronization was
destroyed for all coupling strengths. The only recovery of any kind of
synchronization was achieved by considering a time lag in the
synchronization measures equal to the optical propagation time through
a fiber of the length of the mismatch and only by taking $c\to
1$. This is valuable news, however, as gross length mismatch is
relatively user-controllable and in principle, should be
avoidable. This means that the length of the fiber loop could be a
very effective parameter that would need to be correct in order to
have the ability to recover the message.

The effect of noise in the communications channel was examined
next. An interesting effect was discovered where the lasers actually
synchronized better for lower coupling since less of the optical field
is injected in the cases of weaker coupling, the quality of the
synchronization was effected less. This indicated that a sort of
nonlinear noise reduction was occurring via the coupling scheme, and
hinted towards some possible applications in regards to multi-user
communications.

We ending the section with the presentation of an alternate method of
coupling the lasers. This method was created in an attempt to bypass
all of the synchronization problems found regarding the optical field
phase and state of polarization mismatches. The method uses
electro-optic modulation of the receiver ring laser in an attempt to
establish generalized intensity synchronization alone. Although there
are many unanswered questions regarding the feasibility of the
technique, the results here have been most
encouraging. These issues will be further addressed in~\cite{synchmod}.

In the next section, we turned to an examination of using the above
synchronization behavior insight to attempt to communicate a digital
bit stream utilizing the synchronized pair of ring lasers. A simple ASK
modulation scheme was used to modulate the message onto the intensity of
the chaotic waveform in the transmitter to be recovered at the
receiver via division of the incoming modulated intensity from the
transmitter by the unmodulated receiver laser's intensity.
First we used the pair of identical lasers coupled by direct optical 
coupling, and a remarkable bit error rate was achieved. There was 
error-free recovery of bits ($10^7$ bits were sent) down to a coupling
strength of $c_{crit}=5.0 \times 10^{-3}$. When noise was added to the 
communications channel, the BER was smaller for a range of smaller
coupling strengths. The effect was so prominent that for channel noise
of 60 dB, we obtained error-free recovery in the range of coupling values
$0.1 \le c \le 0.4$ while as $c\to1$ the BER approached $10^{-2}$.

An ASK modulation scheme was also used with a pair of DFRLs coupled
via the alternative coupling-by-intensity modulation method. Again,
remarkable recovery was obtained with error-free recovery down to
a coupling strength of $4.5 \times 10^{-2}$. Further investigation
on the robustness of this method will appear in~\cite{synchmod}.

Finally, we provided a cryptographic context to the communications strategies discussed by us. This context goes well beyond its appearance in optical systems~\cite{sush}. While we do not provide a cryptographic analysis of our suggested communications methods, we have tried to make clear what issues one must address in making any claim that such methods have cryptographic security. From our limited reading of the cryptographic literature on self-synchronous stream ciphers~\cite{ruepp}, it is clear that some nonlinear functions may have security and others certainly will not. Any claim of security should be backed up by an appropriate cryptanalysis and not based in a subjective sense that complex waveforms are secure. Indeed, it seems to us that the interest in chaotic communication, optical or otherwise, is not founded in security of communications but in other advantages of nonlinear systems.

There are several directions in which one may pursue the work reported here. A straightforward possibility is the consideration of other rare-earth-doped fibers where $T_1$ is shorter. With Pr or Nd one can achieve $T_1$'s as small as 100 $\mu\,$s, and this would change many of the features we have reported. It is likely that the sensitivity of synchronization to polarization or phase mismatches would remain, but while sacrificing synchronization for such small values of $c$, we may accomplish other goals such as smaller bit error rates in the presence of channel noise associated with larger conditional Lyapunov exponents on the synchronization manifold. Another direction would be to replace the active element in the ring lasers with other devices, and semiconductor lasers immediately suggest themselves. With these $T_1 \approx 1\,$ns, and many of the operating characteristics investigated here change. We shall report on an investigation of this class of chaotic transmitter and receiver~\cite{semi}.

\section*{Acknowledgments}
We thank the members of INLS, Alistair Mees, and Ulrich Parlitz for helpful discussions
on this subject. This work was part of a joint UCSD/Georgia Tech/Cornell
effort, and we are grateful to Steve Strogatz, Raj Roy, and Govind Agrawal and others in
that program for detailed discussion of the issues here. This work was
supported in part by the U.S. Department of Energy, Office of Basic
Energy Sciences,
Division of Engineering and Geosciences, under grant
DE-FG03-90ER14138,
in part by National Science Foundation grant NCR-9612250, and
in part by the Army Research Office, DAAG55-98-1-0269, MURI Project in Chaotic Communication.

\clearpage

\clearpage

\clearpage

\clearpage

\clearpage

\clearpage

\clearpage

\clearpage

\clearpage

\clearpage

\clearpage

\clearpage

\clearpage

\begin{table}

\caption{Typical parameters for the EDFRL model simulations}
\begin{tabular}{rcc}
Quantity & Symbol & Value \\
\tableline
Linear birefringence & $\Delta$ & $1.8 \times 10^{-6}$ \\
Fiber index of refraction & $n_0$ & $1.45$ \\
External injection amplitude & $A$ & 0.0 \\
Pump strength & $Q$ & $2.4 \times 10^{-2}$ \\ 
Overall gain term & $G$ & $1.35 \times 10^{-2}$ \\
Absorption coefficients & $R_x$, $R_y$ & $0.45$, $0.449995$ \\
Polarization controller angles & $\theta_1$, $\theta_2$, $\theta_3$ & $0.5$, $1.2$, $1.5$ \\
Nonlinear phase shift & $\Phi_{nl}$ & $1.5 \times 10^{-2}$ \\
Round trip time & $\tau_R$ & $200$ ns \\
Excited state lifetime & $T_1$ & $10$ ms \\
Polarization dephasing time & $T_2$ & $1$ ps \\
GVD coefficient & $\beta_2$ & $-20 \mathrm{ ps}^2/\mathrm{km}$ \\
Active fiber length & $l_A$ & $20$ m \\
Passive fiber length & $l_F$ & $20$ m \\
\end{tabular}
\label{tab:parameters}
\end {table}

\end{document}